\documentclass[11pt]{article}
\usepackage{amsmath, amsfonts, setspace, url, verbatim, xcolor, cite, graphicx, todonotes, cancel, slashed, mathtools}
\usepackage[T1]{fontenc}
\usepackage[utf8]{inputenc}
\usepackage[
colorlinks=true, 
linkcolor=vub,   
citecolor=vub,   
filecolor=magenta, 
urlcolor=cyan,
]
{hyperref}
\usepackage[a4paper,top=2cm,left=2.5cm,right=2.5cm,bottom=2cm]{geometry}
\usepackage{microtype}
\newcommand{\be}{\begin{equation}}
\newcommand{\ee}{\end{equation}}
\numberwithin{equation}{section}
\newcommand{\dd}{\text{d}}
\newcommand{\fM}{\mathcal{M}}
\newcommand{\fN}{\mathcal{N}}

\newcommand{\fK}{\mathcal{K}}
\newcommand{\fL}{\mathcal{L}}
\newcommand{\fA}{\mathcal{A}}
\newcommand{\fB}{\mathcal{B}}

\renewcommand{\AA}{{\mathsf{A}}}
\newcommand{\BB}{{\mathsf{B}}}
\newcommand{\CC}{{\mathsf{C}}}
\newcommand{\DD}{{\mathsf{D}}}
\newcommand{\EE}{{\mathsf{E}}}
\newcommand{\FF}{{\mathsf{F}}}

\newcommand{\gM}{\mathcal{M}}
\newcommand{\cH}{\mathcal{H}}

\newcommand{\Hdef}{\mathbf{H}} 
\newcommand{\hdef}{\mathbf{h}} 
\newcommand{\Cdef}{\mathbf{C}}
\newcommand{\Fdef}{\mathbf{F}}

\newcommand{\Aa}{\mathcal{A}}

\newcommand{\Fa}{\mathcal{F}}
\newcommand{\Fb}{\mathcal{H}}
\newcommand{\Fc}{\mathcal{J}}

\newcommand{\Gfour}{\mathrm{SL}(5)}

\newcommand{\Gsix}{\mathrm{E}_{6(6)}}

\newcommand{\Edd}{\mathrm{E}_{d(d)}}

\newcommand{\Odd}{\mathrm{O}(d,d)}

\newcommand{\gendil}{\mathbf{d}}
\definecolor{vub}{RGB}{0,52,154}
\definecolor{vubo}{RGB}{255,102,0}
\definecolor{redd}{RGB}{255,40,40}
\definecolor{r}{RGB}{228,32,20}
\definecolor{o}{RGB}{238,69,4}
\definecolor{y}{RGB}{253,228,1}
\definecolor{g}{RGB}{108,160,0}
\definecolor{b}{RGB}{0,162,203}
\definecolor{i}{RGB}{120,42,117}

\newcommand{\rpm}{+}
\newcommand{\rmp}{-}
\newcommand{\UEDA}{({E}_{\text{EDA}}){}}
\newcommand{\Fthree}{h}
\newcommand{\tFfour}{\widetilde{h}}

\def\T5{T_\text{M5}}

\newcommand{\ii}{\mathsf{i}}
\newcommand{\jj}{\mathsf{j}}
\newcommand{\kk}{\mathsf{k}}
\renewcommand{\ll}{\mathsf{l}}

\newcommand{\MM}{\mathsf{M}}
\newcommand{\NN}{\mathsf{N}}
\newcommand{\KK}{\mathsf{K}}
\newcommand{\LL}{\mathsf{L}}

\newcommand{\GG}{\mathcal{G}}
\newcommand{\hI}{\mathsf{I}}
\newcommand{\hJ}{\mathsf{J}}
\newcommand{\hK}{\mathsf{K}}

\newcommand{\as}{\mathsf{a}}
\newcommand{\bs}{\mathsf{b}}
\newcommand{\cs}{\mathsf{c}}
\newcommand{\ds}{\mathsf{d}}

\newcommand{\Mfour}{\mathfrak{M}^{\tilde f}_4}
\newcommand{\Mone}{\mathfrak{M}_d}
\newcommand{\Mtwo}{\widetilde{\mathfrak{M}}_{\tilde d}}
\newlength{\bibitemsep}
\newlength{\bibparskip}
\let\oldthebibliography\thebibliography
\renewcommand\thebibliography[1]{%
\oldthebibliography{#1}%
\setlength{\parskip}{\bibitemsep}%
\setlength{\itemsep}{\bibparskip}%
}

\begin{document}

\begin{center}

{\Large \bf
Non-isometric U-dualities
}
\vskip 2em
{\large \bf  Chris D. A. Blair}
\vskip 1em
{\it  
Theoretische Natuurkunde, Vrije Universiteit Brussel, and the International Solvay Institutes, \\ Pleinlaan 2, B-1050 Brussels, Belgium 
\\ {\tt Christopher.Blair@vub.be}
}
\vskip 0em 
\end{center}

\begin{abstract}
I study generalisations of U-duality transformations which do not rely on the existence of isometries.
I start by providing more details of a recently proposed generalised U-duality map between solutions of type IIA supergravity of the form $\text{M}_7 \times \text{S}^3$, with NSNS flux, and solutions of 11-dimensional supergravity, in which the three-sphere is replaced by a four-dimensional geometry which encodes three-algebra structure constants. 
I then show that when $\text{M}_7$ admits two abelian isometries, TsT deformations on the IIA side become six-vector deformations in the 11-dimensional setting.
These six-vector deformations involve an action of $\mathrm{E}_{6(6)}$ on both isometric and non-isometric directions. 
I discuss the algebraic interpretation of these deformations, and compare and contrast them with (generalised) Yang-Baxter deformations in supergravity.

\vspace{1em}

\end{abstract}


{
\hypersetup{linkcolor=vub}
\tableofcontents
}

\clearpage

\section{Introduction}
\label{intro}

Supergravity inherits T- and U-duality symmetry from string and M-theory.
These dualities can be used to map solutions with abelian isometries to other solutions with abelian isometries.
Broader classes of transformations between solutions with more complicated or even no isometries can be considered to be \emph{generalised dualities}.
The most studied examples are generalised T-dualities such as non-abelian \cite{delaOssa:1992vci} and Poisson-Lie \cite{Klimcik:1995dy,Klimcik:1995ux} T-duality.
These have well-established applications as solution generating mechanisms in supergravity and close links to integrable deformations such as the Yang-Baxter deformation \cite{Klimcik:2002zj}.
More recently, thanks in large part to progress in understanding the generalised geometric/double field theory interpretation of these generalised T-dualities \cite{Hassler:2017yza,Demulder:2018lmj,Sakatani:2019jgu,Catal-Ozer:2019hxw}, progress has been made extrapolating features of generalised dualities to the case of U-duality \cite{Sakatani:2019zrs,Malek:2019xrf,Sakatani:2020iad, Blair:2020ndg,Malek:2020hpo,Sakatani:2020wah,Musaev:2020bwm,Musaev:2020nrt,Bugden:2021wxg,Blair:2022gsx}.
The motivation for this paper is to further study the realisation of new non-isometric U-dualities in 11 dimensions.

The following guideline or definition for a generalised duality in supergravity can be extracted from the known examples.
This definition is simply the existence of a map between solutions of supergravity theory 1, with spacetime $\text{M}_{n} \times \Mone$, to solutions of supergravity theory 2, with spacetime $\text{M}_n \times \Mtwo$, i.e.
\be
\begin{array}{ccc}
\textbf{Theory 1} & & \textbf{Theory 2}\\
\text{M}_{n} \times \Mone & \longleftrightarrow &\text{M}_{n} \times \Mtwo \,,\\
\end{array}
\label{generalMap}
\ee
such that reducing theory 1 on the $d$-dimensional geometry $\Mone$ gives the same lower-dimensional theory as reducing theory 2 on the $\tilde d$-dimensional geometry $\Mtwo$.\footnote{I denote the dimensions as $d$ and $\tilde d$ to allow for duality between 10- and 11-dimensional theories.}
The appropriate sense of dimensional reduction required is that of a consistent truncation, so that solutions of the common lower dimensional theory can be lifted to solutions of either theory 1 and theory 2 in higher dimensions. 
These uplifted solutions are then regarded as `dual', at the very least providing a solution generating mechanism in supergravity.

In this paper, I study an example of this type of map, developed in \cite{Blair:2020ndg,Blair:2022gsx}, of the form:
\be
\begin{array}{ccc}
\textbf{IIA} & & \textbf{M-theory}\\
\text{M}_7 \times \mathrm{S}^3& \longleftrightarrow &\text{M}_7 \times \Mfour\\
\text{with NSNS flux}&&\text{with `3-algebra flux'}
\end{array} 
\label{genUmap}
\ee
This can be used as a solution generating mechanism in the following way.
Start with the geometries on the left-hand side. 
These are solutions of ten-dimensional type IIA supergravity of the form $\text{M}_7  \times \mathrm{S}^3$, with NSNS flux. These consistently truncate to 7-dimensional $\mathrm{CSO}(4,0,1)$ gauged supergravity \cite{Cvetic:2000dm,Cvetic:2000ah}.
Then (in principle) new solutions of 11-dimensional supergravity are constructed (locally) on the right-hand side by uplifting using the alternative consistent truncation described in \cite{Blair:2020ndg,Blair:2022gsx}.

These new solutions are linked to an underlying non-Lie algebraic structure, which is that of a three-algebra. 
By this I mean an algebra with a totally antisymmetric bracket involving three, rather than two, elements, such that generators $t^a$ obey $[t^a,t^b,t^c] = \tilde f^{abc}{}_d t^d$. The map \eqref{genUmap} involves the four-dimensional case with $\tilde f^{abc}{}_d = \epsilon^{abce} \delta_{ed}$, and it is these structure constants which control the form of the new four-dimensional geometry $\Mfour$. (Despite this, the role, if any, of an actual three-algebra symmetry is unclear, and will make no further appearance in this paper.)
This can be viewed as an M-theoretic generalisation of the sort of solutions which appear after applying non-abelian T-duality \cite{delaOssa:1992vci, Sfetsos:2010uq}, in which the dual geometry is locally determined by the existence of an underlying Lie algebra symmetry.

A proof of concept of the map \eqref{genUmap} has been demonstrated in \cite{Blair:2022gsx}, where the F1-(near horizon) NS5 brane solution of type IIA was taken as the initial solution on the left-hand side of \eqref{genUmap} and used to generate a novel 11-dimensional solution.
However, explicit general formulae applicable to arbitrary backgrounds of the form $\text{M}_7 \times \text{S}^3$ were not presented.
The first result of this paper is to improve this situation, by providing the complete expressions starting with purely NSNS solutions of type IIA.
These are given in section \ref{generalmap}.

I then investigate and explain an interesting feature noticed in the new solution obtained in \cite{Blair:2022gsx}.
In that case, the F1 near horizon limit of the initial IIA solution was of the form $\text{M}_7  = \text{AdS}_3 \times \mathrm{T}_4$.
The corresponding new 11-dimensional solution in the same limit then had the (local) form $\text{AdS}_3 \times \mathrm{T}_4 \times \Mfour$.\footnote{
The solution in this limit could be identified as belonging to a class of M-theory $\text{AdS}_3$ solutions obtained in \cite{Lozano:2020bxo}.
These solutions admit holographic duals similar to the holographic duals of solutions obtained by non-abelian T-duality, again suggesting that these three-algebra geometries are 11-dimensional analogues of such solutions, and also giving one way to globally complete the solution. Conventional U-duality deformations of the solutions of \cite{Lozano:2020bxo} were studied in \cite{Zacarias:2021pfz}.}
The interpolation away from the F1 near horizon region, which can be realised as a TsT transformation, or bivector deformation, was shown in \cite{Blair:2022gsx} to become an $\Gsix$-valued \emph{six-vector deformation} of the AdS limit.
This and other examples of deformations of $\text{AdS}_3 \times \mathrm{T}_4$ are reviewed in section \ref{maptst}.

While the original TsT deformation involves the $\mathrm{O}(2,2)$ T-duality group acting in two isometric directions, the deformation appearing in the new 11-dimensional solution involves an action of $\Gsix$ in six directions.
Two of these correspond to the original isometries of the type IIA, and the remaining four corresponded to the new non-trivial internal geometry and hence to \emph{non-isometric} directions.
The main goal of this paper is to provide a general explanation of how this comes about. This can be found in section \ref{tstgeneral}.

I then discuss, in section \ref{algint}, the algebraic interpretation of this non-isometric duality.
The full underlying algebraic structure behind the right-hand side of \eqref{genUmap} is the Poisson-Lie (or Nambu-Lie) U-duality construction of \cite{Sakatani:2019zrs,Malek:2019xrf}.
This introduces an underlying four-dimensional Nambu-Lie group equipped with a trivector, $\pi^{abc}$. In the case \eqref{genUmap}, this group is abelian and the trivector obeys $L_{v_d} \pi^{abc} = \tilde f^{abc}{}_d$, where $v_a$ are the left-invariant vectors of the group.
The geometry of $\Mfour$ is then determined via the Nambu-Lie group data (left-invariant vectors and forms, and the trivector) using a generalised frame in $\Gfour$ generalised geometry/exceptional field theory. 
This frame generates a generalised parallelisation realising an algebra dubbed the exceptional Drinfeld algebra.

The six-vector parameter deforming the 11-dimensional solutions on the right-hand side of \eqref{genUmap} is:
\be
\Omega^{(6)} 
= \lambda \epsilon^{\as_1 \dots \as_6} V_{\as_1}{}
\dots V_{\as_6}{}
\,, 
\quad V_{\as}  = ( k_{\alpha} , v_a )  \,,
\label{omega6intro}
\ee
where $\as=1,\dots,6$ labels a set of six vectors.
This set comprises the two commuting Killing vectors $k_{\alpha}$ inherited from the isometries of the original type IIA background, as well as the four vectors $v_a$ associated with the underlying Nambu-Lie group structure.
These are \emph{not} Killing (their action on the spacetime fields is determined in terms of the three-algebra structure constants linked to $\Mfour$ \cite{Sakatani:2020iad}).
The fact that this deformation \eqref{omega6intro} nonetheless generates a solution can be linked to the fact that it can be viewed as a transformation leaving invariant the underlying exceptional Drinfeld algebra. 

This can be compared and contrasted with existing approaches to polyvector deformations.
Inspired by the recasting of the Yang-Baxter deformation \cite{Klimcik:2002zj,Klimcik:2008eq,Delduc:2013fga,Matsumoto:2015jja} in terms of a non-constant bivector parameter \cite{Araujo:2017jkb,Araujo:2017jap,Sakamoto:2017cpu,Bakhmatov:2017joy,Sakamoto:2018krs,Borsato:2018idb,Catal-Ozer:2019tmm}, polyvector deformations involving $p$ Killing vectors, of the form $r^{a_1 \dots a_p} k_{a_1} \dots k_{a_p}$, have been studied in both 10- and 11-dimensions \cite{Bakhmatov:2019dow,Sakatani:2019zrs,Malek:2019xrf,Bakhmatov:2020kul,Malek:2020hpo, Gubarev:2020ydf}.
The transformation \eqref{omega6intro} further generalises these generalised Yang-Baxter deformations by dropping the assumption that all vectors involved are Killing. 
This is in fact implicitly what has been used in identifying Yang-Baxter deformed backgrounds and non-abelian T-duals as T-folds, as in \cite{Fernandez-Melgarejo:2017oyu, Bugden:2019vlj}.
There the background is patched by a global T-duality transformation which does not act on isometric directions.
This generalises to the U-fold interpretation of the 11-dimensional solution appearing on the right-hand side of \eqref{genUmap}.
These examples suggest the existence of a broader landscape of polyvector deformations applicable in backgrounds admitting underlying (but non-isometric) Poisson-Lie or Nambu-Lie symmetry.

The complete outline of this paper is as follows:

I first discuss in section \ref{genmetpoly} the necessary technical language surrounding generalised metrics, polyvectors and generalised duality.

In section \ref{generalmap}, I discuss the explicit form of the map \eqref{genUmap} in more detail.
In section \ref{maptst}, I motivate the investigation of six-vector transformations with examples where $\text{M}_7 = \text{AdS}_3 \times \mathrm{T}^4$, and then give the general explanation of how TsT deformations are transmogrified into $\Gsix$ six-vector deformations in section \ref{tstgeneral}.
I further briefly discuss here the fate of more general T-duality transformations, which can also be interpreted as non-isometric $\Gsix$ transformations generically with a less clear geometric interpretation than the six-vector deformation.

In section \ref{algint}, I discuss the algebraic interpretation of the six-vector deformation, and its similarities to, and differences with, other sorts of polyvector deformations.

In section \ref{discussion}, I conclude with a discussion.

Appendix \ref{conventions} describes my conventions relating to forms and to generalised geometry.

Appendix \ref{appredup} describes the derivation of the map \eqref{genUmap} in detail.

Appendix \ref{genkill} highlights a non-trivial self T-duality property of $\text{AdS}_3 \times \text{T}^4 \times \text{S}^3$ with NSNS flux.

\section{Generalised geometry and duality}
\label{genmetpoly} 

\subsection{Generalised duality from generalised geometry} 

A convenient way to formulate generalised duality as a map of the form \eqref{generalMap} is to use double or exceptional geometric approaches \cite{Hassler:2017yza,Demulder:2018lmj,Sakatani:2019jgu,Catal-Ozer:2019hxw,Sakatani:2019zrs,Malek:2019xrf
}, where supergravity is reformulated -- in arbitrary backgrounds -- in terms of variables on which $\Odd$ or $\Edd$ transformations act linearly \cite{Gualtieri:2003dx,Hull:2007zu}, see \cite{Musaev:2019zcr,Berman:2020tqn} for reviews. 
The starting point is to regard the spacetime geometry as having the form $\text{M}_n \times \Mone$ and work on the generalised tangent bundle of $\Mone$.
In the simplest cases, this bundle is $\mathcal{E} \approx T \Mone \oplus \Lambda^{n} T^* \Mone$, pairing vectors and $n$-forms into `generalised vectors'. 

The metric, $G$, and $p$-form gauge fields, collectively denoted $C$, on $\Mone$ are combined into a \emph{generalised metric} $\gM(G,C)$ on the generalised tangent bundle.
This is itself an $\Odd$ or $\Edd$ group element, and transforms as $\gM \rightarrow \widetilde{\gM} = \mathcal{O} \gM \mathcal{O}^T$ under $\Odd$ or $\Edd$ duality transformations.
If $\Mone$ has $d$ abelian isometries, this automatically realises T- or U-duality.

To describe generalised duality of more general backgrounds, a little more detail is needed.
First of all, it is required that there exists a factorisation 
\be
\gM_{MN} (G,C) = \Delta^{-1} E_M{}^A E_N{}^B  M_{AB}  
\label{Mfactor}
\ee
in which the generalised vielbein $E^A$ and the scalar $\Delta$ depend only on the coordinates on $\Mone$, while $M_{AB}$ is independent of these coordinates.
Under certain differential conditions on $E^A$, this permits a dimensional reduction to a lower dimensional gauged maximal supergravity.
These differential conditions have a very simple interpretation using the generalised Lie derivative (see equation \eqref{genLie}) of generalised geometry, and in terms of the inverse generalised frame $E_A$ have the form 
\be
\mathcal{L}_{E_A} E_B = -F_{AB}{}^C E_C\,,\quad 
\mathcal{L}_{E_A} \Delta = \theta_A \Delta \,.
\label{genpar} 
\ee
for constants $F_{AB}{}^C$ and $\theta_A$. 
This defines a `generalised parallelisation' and permits a consistent truncation where $F_{AB}{}^C$ and $\theta_A$ become the gaugings of the lower-dimensional gauged supergravity. (The case of interest to this paper has $\theta_A=0$.)

The map \eqref{generalMap} is possible when there exists a different background $\Mtwo$ allowing for a generalised vielbein $\widetilde{E}^A$ obeying the same differential constraints and permitting an alternative consistent truncation to lower dimensions.
Then both theory 1 and theory 2 reduce, on $\Mone$ and $\Mtwo$ respectively, to a common lower dimensional theory, and solutions of this theory (given here by a specific $M_{AB}$) can be uplifted to solutions of either theory 1 or theory 2.

If $\gM(G,C) =\Delta^{-1} E^A M_{AB} E^B$ describes the geometry of the first background, the generalised metric $\widetilde{\gM}(\widetilde G, \widetilde C) = \widetilde \Delta^{-1} \widetilde{E}^A M_{AB} \widetilde E^B$ will describe the geometry of the second.
This implicitly describes a map between $G,C$ and $\widetilde{G}, \widetilde{C}$, which evidently depends on the form of the generalised vielbeins involved, and which takes solutions to solutions.\footnote{This could also somewhat radically be viewed as a coordinate dependent $\Odd$ or $\Edd$ transformation simultaneously involving the coordinates on $\Mone$ and $\Mtwo$.}

In analysis of generalised dualities, a special role is played by the algebra defined by \eqref{genpar}, with structure constants $F_{AB}{}^C$. 
Introduce generators $T_A$ obeying $[T_A, T_B] = F_{AB}{}^C T_C$. In the exceptional case, this may not be an antisymmetric bracket.
The existence of a particular generalised parallelisation on the generalised tangent bundle of $\Mone$ or $\Mtwo$ can be linked to the structure of this algebra. 

For instance, in the Poisson-Lie cases, a selection is made of $d$ `physical' generators $T_a$ corresponding to an isotropic subalgebra, obeying $[T_a, T_b]= f_{ab}{}^c T_c$ and $\langle T_a, T_b \rangle = 0$ for a particular bilinear bracket (corresponding to the $\Odd$ invariant bilinear form in the T-duality case, and to a symmetric intertwining map involving $\Edd$ representations in the U-duality case).
The corresponding frame fields $E_a$ can then be constructed out of the left-invariant vector fields of a Poisson-Lie (or Nambu-Lie) group for which $f_{ab}{}^c$ are the Lie algebra structure constants.
The remaining differential equations in \eqref{genpar} are further solvable thanks to the existence of additional structure on the group in the form of a polyvector (or polyvectors).
This construction is used on the right-hand side of \eqref{genUmap}.
The left-hand side is not \emph{strictly} in the same category, as the choice of physical generators although isotropic does not give a subalgebra due to the NSNS flux (instead $[T_{\tilde a}, T_{\tilde b}] = H_{\tilde a \tilde b \tilde c} T^{\tilde c}$, $\tilde a=1,2,3$, where $T^{\tilde a}$ are other generators of the full algebra).\footnote{I say strictly, but at least in the T-duality case it is known that the Poisson-Lie framework can be generalised to allow for the presence of H-flux and an isotropic non-subalgebra \cite{Klimcik:2015gba}, see \cite{Demulder:2018lmj} for a generalised geometric discussion, and it is likely that the U-duality case admits a similar extension allowing a unified algebraic approach, as noted already in \cite{Blair:2020ndg}.}
While a generalised geometric frame is known realising this consistent truncation \cite{Lee:2014mla, Hohm:2014qga}, the original formulae of \cite{Cvetic:2000dm,Cvetic:2000ah} in fact suffices to formulate \eqref{genUmap} explicitly, as used in \cite{Blair:2022gsx} (and here in appendix \ref{appredup}).

\subsection{Generalised metrics and polyvectors}

I will frequently implement duality transformations, of various types, in terms of an action of $\Odd$ or $\Edd$ on the natural `duality covariant' fields, such as the generalised metric.
I provide here a brief review of the basic structure of generalised metrics, and of the $\Odd$ or $\Edd$ description of polyvector transformations.
Some of the underlying logic behind my conventions can be found in appendix \ref{conventions}.

Multiple uses of polyvectors will feature in this paper.
Firstly, when $\Mone$ admits abelian isometries, the $\Odd$ or $\Edd$ transformations generated by constant bi- or polyvectors constitute non-trivial non-geometric duality transformations.
In the $\Odd$ case, this corresponds to TsT transformations \cite{Lunin:2005jy,Catal-Ozer:2005dux}.
Secondly, coordinate dependent polyvectors will be used when formulating generalised vielbeins meeting the conditions leading to the existence of the generalised duality map \eqref{generalMap}.
Lastly, a constant $\Gsix$ polyvector will appear when analysing the fate of TsT deformations after generalised duality.
This will involve $\Gsix$ acting on a combination of isometric and non-isometric directions.
I now review the background needed to understand each of these cases.

\subsubsection*{$\Odd$ T-duality and bivectors}

In the T-duality case, the generalised tangent bundle is $\mathcal{E} \approx T\Mone \oplus T^* \Mone$.
A generalised vector is a pair $V=(v,\omega_{(1)})$ of a vector $v$ and one-form $\omega_{(1)}$.
I introduce a metric $G$ and two-form $B$ on $\Mone$.
These are encoded in a generalised metric defined on the generalised tangent bundle:
\be
\mathcal{H} = \begin{pmatrix} 1 & B \\ 0 & 1 \end{pmatrix} 
\begin{pmatrix} G & 0 \\ 0 & G^{-1} \end{pmatrix}
\begin{pmatrix}
1 & 0 \\
- B& 1 
\end{pmatrix} \,.
\label{Hinit}
\ee
The matrix involving the two-form $B$ is itself an $\Odd$ element.
I can introduce a bivector $\beta$ in the form of the `transposed' transformation:
\be
U_B \equiv 
\begin{pmatrix} 
1 & B \\
0 & 1
\end{pmatrix}
\leftrightarrow 
\begin{pmatrix} 
1 & 0 \\
\beta & 1
\end{pmatrix} \equiv U_\beta \,,
\label{defUbeta}
\ee
I will need explicit expressions for the action of such a transformation applied to the NSNS sector of type II supergravity.
The fields here are the string frame metric, $G$, the Kalb-Ramond two-form, $B$, and the dilaton, $\varphi$.
I take the background geometry to have the form $M_{10-d} \times \Mone$.
Sometimes it is convenient to take $d=10$ (and so describe the whole spacetime in unified fashion) or to take $d<10$ (for instance to restrict to the part of the spacetime which actually admits isometries and hence can be T-dualised).
In either case, the $\Odd$ generalised metric, $\cH_{MN}$, is parametrised in terms of the metric and $B$-field on $\Mone$ as in \eqref{Hinit}.
The action of a transformation $\mathcal{O} \in \Odd$ is:
\be
\mathcal{H}_{MN} \rightarrow \widetilde{\mathcal{H}}_{MN} = \mathcal{O}_M{}^K \mathcal{H}_{KL} \mathcal{O}_N{}^L \,.
\label{oddaction}
\ee
I will realise TsT transformations, or bivector shifts, by acting with $\mathcal{O} = U_\beta$, with a constant bivector.
In my conventions, a bivector with non-zero components $\beta^{12} = - \beta^{21} = \lambda$ corresponds to the combination of: T-duality in $x^2$, shift $x^1 \rightarrow x^1 + \lambda \tilde x^2$, T-duality back on $\tilde x^2$. Alternatively: T-dualise in $x^1$ and $x^2$, shift the B-field as $B_{12} \rightarrow B_{12} +\lambda$ and T-dualise back on $\tilde x^1$ and $\tilde x^2$.

I decompose the 10-dimensional metric and B-field as:
\be
\begin{split}
\dd s^2_{10} & = G_{\mu\nu} \dd x^\mu \dd x^\nu + G_{\alpha \beta} Dx^\alpha Dx^\beta \,,\\
B & = \tfrac12 (B_{\mu\nu} + A_{\mu}{}^\alpha A_{\nu \alpha} ) \dd x^\mu \wedge \dd x^\nu 
+ A_{\mu\alpha} \dd x^\mu \wedge D x^\alpha + \tfrac12 B_{\alpha \beta} D x^\alpha \wedge D x^\beta \,,
\end{split} 
\label{10metricdec}
\ee
where $D x^\alpha \equiv \dd x^\alpha + A_\mu{}^\alpha \dd x^\mu$, with $\alpha,\beta=1,\dots ,d$ and $\mu,\nu=1,\dots, 10-d$.
The components of the $d$-dimensional metric, $G_{\alpha\beta}$, and $B$-field, $B_{\alpha \beta}$, enter into the $\Odd$ generalised metric as in \eqref{Hinit}.
The pair $(A_\mu{}^\alpha, A_{\mu \alpha})$ transform as an $\Odd$ vector.
Both $G_{\mu\nu}$ and $B_{\mu\nu}$ are invariant, as is the $\Odd$ invariant generalised dilaton, $\gendil$, which is defined by $e^{-2\gendil} = e^{-2\varphi} \sqrt{|G|}$ in terms of the determinant of the metric on $\Mone$.
I also assemble the field strengths, defining
\be
F_{(2)}^\alpha = \dd A_{(1)}^\alpha \,,\quad 
 H = \Hdef_{(3)} + \Hdef_{(2) \alpha} \wedge D x^\alpha + \Hdef_{(1) \alpha\beta} \wedge D x^\alpha\wedge D x^\beta \,,
\label{TFs}
\ee
where $H=\dd B$ is the original field strength for the 10-dimensional 2-form.
Here I assume explicitly that $x^\alpha$ are adapted coordinates for isometries, and use the subscript $(p)$ to denote a $(10-d)$-dimensional $p$-form.
Under $\Odd$, $\Hdef_{(3)}$ is invariant while
\be
\mathcal{F}_{(2)}^M = \begin{pmatrix} 
F_{(2)}^\alpha \\
\Hdef_{(2)\alpha}+ B_{\alpha \beta} F_{(2)}^\beta 
\end{pmatrix} 
\label{TF}
\ee
transforms as a vector.

I now specialise to the case $d=2$.
I can let $B_{\alpha \beta} = B \epsilon_{\alpha \beta}$.\footnote{I take $\epsilon_{12} = \epsilon^{12} = 1$ so that $\epsilon^{\alpha \gamma} \epsilon_{\beta \gamma} =\delta^\alpha_\beta$.}
Under a bivector transformation with $\beta^{\alpha \beta} = \lambda \epsilon^{\alpha \beta}$, 
\be
\begin{split} 
\widetilde{G}_{\alpha \beta} & = (1-2 \lambda B + \lambda^2 ( \det G_{\alpha \beta} + B^2 ) )^{-1}G_{\alpha \beta} \,,\\
\widetilde{B} & = (B-\lambda(\det G_{\alpha\beta} +B^2))(1-2 \lambda B + \lambda^2 ( \det G_{\alpha \beta} + B^2 ) )^{-1} \,,\\
e^{-2\tilde \varphi} & = (1-2 \lambda B + \lambda^2 ( \det G_{\alpha \beta} + B^2 ) ) e^{-2\varphi} \,,\\
\widetilde{A}_\mu{}^{\alpha} &= A_\mu{}^\alpha + \lambda \epsilon^{\alpha \beta} A_{\mu \beta} \,,
\label{bivrules}
\end{split}
\ee
with the other fields invariant.
 
Later on, I will need to use the following factorisation of the $\mathrm{O}(2,2)$ generalised metric.
Define $\bar{\cH}$ by $\bar{\cH}_{\alpha\beta} = \cH_{\alpha \beta}$, $\bar{\cH}_{\alpha \bar \beta} = \bar{\cH}_{\bar \beta \alpha} = \epsilon_{\beta \gamma} \cH_{\alpha}{}^\gamma$, $\bar{\cH}_{\bar \alpha \bar \beta} = \epsilon_{\alpha \gamma} \epsilon_{\beta \delta} \cH^{\gamma \delta}$.
Then $\bar{\cH} = \mathcal{H}_{\tau} \otimes \mathcal{H}_{\rho}$ with
\be
\mathcal{H}_{\tau} = |\det G_{\alpha \beta}|^{-1/2} G_{\alpha \beta} \,,\quad
\mathcal{H}_{\rho} = |\det G_{\alpha \beta}|^{1/2} 
\begin{pmatrix}
1+ (\det G_{\alpha\beta})^{-1} B^2 & (\det G_{\alpha\beta})^{-1} B 
\\
(\det G_{\alpha\beta})^{-1} B  & (\det G_{\alpha\beta})^{-1}
\end{pmatrix} \,.
\label{Htaurho} 
\ee
This realises the identification $\mathrm{O}(2,2) \sim \mathrm{SL}(2)_\tau \times \mathrm{SL}(2)_\rho$ with each of $\mathcal{H}_\tau$ and $\mathcal{H}_\rho$ transforming under separate copies of $\mathrm{SL}(2)$.
Here $\tau=\tau_1+i\tau_2$ is to be viewed as the complex structure of the metric $G_{\alpha\beta}$, while $\rho = B+ i |\det G_{\alpha\beta}|^{1/2}$ encodes the $B$-field and volume modulus.
Transformations in $\mathrm{SL}(2)_\tau$ are part of the geometric subgroup of $\mathrm{O}(2,2)$ corresponding to (volume preserving) global general coordinate transformations.
The transformations in $\mathrm{SL}(2)_\rho$ are more interesting, and include volume rescalings, $B$-shifts, T-duality transformations on both directions (via the $\mathrm{SL}(2)$ inversion element) and bivector shifts, with the latter acting as
\be
\mathcal{H}_{\rho} \rightarrow U_\lambda \mathcal{H}_\rho U_\lambda^T\,,\quad 
U_\lambda = \begin{pmatrix} 1  & 0 \\ - \lambda & 1 \end{pmatrix} \,.
\label{Hrholambda}
\ee
Note also that T-duality on one direction corresponds to a $\mathbb{Z}_2$ transformation swapping $\tau$ and $\rho$.

\subsubsection*{$\Gfour$ (generalised) U-duality and trivectors}

In the U-duality case, with $\text{M}_n \times \Mone$ a background of 11-dimensional supergravity, the generalised tangent bundle is $\mathcal{E} \approx T \Mone \oplus \Lambda^2 T^*\Mone \oplus \Lambda^5 T^* \Mone \oplus \dots$, with additional factors only relevant for $d \geq 7$. 
My conventions for the generalised Lie derivative, and the gauge transformations of the 3- and 6-form that it encodes, are set out in appendix \ref{conventions}.

I consider first the $d=4$ case. 
I introduce a metric $g$ and three-form $C$ on $\Mone$.
The $\Gfour$ generalised metric is: 
\be
\mathcal{M} = |\det g|^{\tfrac{1}{5}}\begin{pmatrix} 1 & C \\ 0 & 1 \end{pmatrix} 
\begin{pmatrix} g & 0 \\ 0 & g^{-1}g^{-1} \end{pmatrix}
\begin{pmatrix}
1 & 0 \\
C& 0 
\end{pmatrix} \,.
\label{Minit}
\ee
The conformal factor ensures that $\det \mathcal{M}=1$, otherwise the generalised metric will not be a true $\Gfour$ element.
As a generalised vector consists of a vector and a two-form, the indices on the term $g^{-1} g^{-1}$ are to be understood as being antisymmetrised, thus $(g^{-1} g^{-1})^{ij,kl} = 2 g^{i[k} g^{l]j}$.
I use a contraction convention where for $V^M = ( v^i, \omega_{ij} )$ and $W_M = ( w_i , \nu^{ij})$ I have $V^M W_M = v^i w_i + \tfrac12 \omega_{ij} \nu^{ij}$.

The matrix involving the three-form is an $\Edd$ transformation, and I introduce a trivector as the transposed transformation:
\be
U_C \equiv
\begin{pmatrix} 
1 & C \\
0 & 1
\end{pmatrix}
\leftrightarrow 
\begin{pmatrix} 
1 & 0 \\
\Omega & 1
\end{pmatrix}\equiv U_\Omega \,.
\ee
A non-constant trivector is used to define the right-hand side of the generalised U-duality map \eqref{genUmap}.
This trivector appears as part of the auxiliary geometric data defining the right-hand side of \eqref{genUmap} via the Poisson-Lie or Nambu-Lie U-duality construction of \cite{Sakatani:2019zrs,Malek:2019xrf}.
This introduces an underlying Nambu-Lie group $\mathfrak{G}$, which in this case is four-dimensional, with 
associated left-invariant vector fields, $v^a$, and forms, $l^a$, obeying $L_{v_a} v_b =  - f_{ab}{}^c v_c$ and $d l^a = \tfrac12 f_{bc}{}^a l^b \wedge l^c$, where $f_{ab}{}^c$ are the structure constants of the Lie algebra $\mathfrak{g}$ of $\mathfrak{G}$. Here $a,b=1,\dots,4$.
In addition, the group is equipped with a trivector, denoted $\pi^{abc}$. 
By definition, this obeys:
\be
d \pi^{abc} = \tilde f^{abc}{}_d l^d + 3 f_{ed}{}^{[a} \pi^{bc]d} l^e + \tfrac13 \pi^{abc} \mathfrak{L}_d l^d \,,
\ee
where $\tilde f^{abc}{}_d$ are structure constants for a three-algebra.
The additional constant $\mathfrak{L}_a$ is identified with the derivative of an additional scalar, $L_{v_a} \ln \alpha = \tfrac13 \mathfrak{L}_a$.
Using the above differential conditions (see \cite{Sakatani:2019zrs,Malek:2019xrf} for full details) the generalised frame defined by
\be
\UEDA_A{}^M  = \begin{pmatrix}
v_a{}^i & 0 \\
\pi^{abc} v_c{}^i & 2\alpha l^{[a}{}_i l^{b]}{}_j 
\end{pmatrix} \,,\quad
\Delta = ( \alpha^3 \det l )^{1/5} \,,
\label{fiveEDAframe} 
\ee
then obeys the algebra \eqref{genpar} with structure constants of the so-called exceptional Drinfeld algebra (EDA).
This algebra is defined by following (not necessarily antisymmetric) brackets:
\be
\begin{split} 
[T_a, T_b]& = f_{ab}{}^c T_c \,,\quad [T^{ab}, T^{cd}] =  2 \tilde f^{ab[c}{}_e T^{d]e} \,, \\ 
[T_a, T^{bc} ]& = 2 f_{ad}{}^{[b} T^{c]d} - \tilde f^{bcd}{}_a T_d - \tfrac13 \mathfrak{L}_a T^{bc} \,,\\
[T^{bc}, T_a] &= 3 f_{[de}^{[b} \delta^{c]}_{a]} T^{de} + \tilde f^{bcd}{}_a T_d + \mathfrak{L}_d \delta_a^{[b} T^{cd]} \,.
\end{split} 
\label{fiveEDA}
\ee
The subalgebra generated by the generators $T_a$ is a Lie algebra. 
This is the distinguished isotropic subalgebra whose selection corresponds to a particular choice of decomposition of the EDA, leading to the construction of the frame \eqref{fiveEDAframe}. 

The right-hand side of the generalised U-duality map \eqref{genUmap} corresponds \cite{Blair:2020ndg} to the case where the group $\mathfrak{G}$ is abelian, so $f_{ab}{}^c=0$, but the three-algebra is take to be the Euclidean three-algebra with $\tilde f^{abc}{}_d = \epsilon^{abce} \delta_{ed}$.\footnote{In addition, $\mathfrak{L}_a=0$. This implies that the `trombone' gauging $\theta_A$ in \eqref{genpar} also vanishes.}
Introducing coordinates $x^i$, $i=1,\dots,4$, the corresponding geometric data are then 
\be
v_a{}^i = \delta_a^i \,,\quad l_i{}^a = \delta_a^i \,,\quad \pi^{abc} = \epsilon^{abcd} x_d\,,\quad
\alpha=1 \,.
\label{moredata}
\ee
The algebra \eqref{fiveEDA} then corresponds to $\mathrm{CSO}(4,0,1)$, which is to say the Euclidean Poincar\'e algebra.
The generators $T_a$ correspond to the abelian translational subalgebra, while $T^{ab}$ generate the rotational $\mathrm{so}(4)$ Lie algebra.\footnote{This is clearest to see by defining dualised generators $T_{ab} = \tfrac12 \epsilon_{abcd} T^{cd}$ and noting that the Euclidean metric $\delta_{ab}$ is encoded in the three-algebra structure constants, again by dualisation with $\epsilon^{abcd}$.}
The generalised frame \eqref{fiveEDAframe} built using \eqref{moredata} then allows for a consistent truncation to 7-dimensional $\mathrm{CSO}(4,0,1)$ gauged supergravity.

To describe this consistent truncation fully, I should introduce the additional $\Gfour$ covariant fields needed to capture all the degrees of freedom of 11-dimensional supergravity.
This is presented in appendix \ref{appredup}.

\subsubsection*{$\Gsix$ U-duality and six-vectors} 

I now discuss the action of $\Gsix$ transformations.
The generalised tangent bundle is $\mathcal{E} \approx T \Mone \oplus \Lambda^2 T^*\Mone \oplus \Lambda^5 T^* \Mone$, with $d=6$.
I will let $\ii,\jj=1,\dots,6$, reserving the indices $i,j$ for four-dimensional use.
The fundamental representation of $\Gsix$ is 27-dimensional.
A generalised vector $V^{\MM}$ in this representation is built using a vector, two-form and five-form, thus $V^{\MM} = ( V^{\ii}, V_{\ii_1\ii_2}, V_{\ii_1 \dots \ii_5})$.
It is often convenient to dualise the five-form index, letting $V^{\bar \ii} = \tfrac{1}{5!} \epsilon^{\ii \jj_1 \dots \jj_5} V_{\jj_1 \dots \jj_5}$.

The fields of 11-dimensional supergravity restricted to $\Mone$ are the metric, which I denote $\phi_{\ii\jj}$, the three-form, $C_{\ii\jj\kk}$, and the dual six-form, $C_{\ii_1 \dots \ii_6}$.
The latter has only 1 independent component, so I write it as $C_{\ii_1 \dots \ii_6} = C_6 \epsilon_{\ii_1 \dots \ii_6}$.
These fields can be encoded in a generalised metric \cite{Berman:2011jh,Coimbra:2011ky}, with the following parametrisation:
\be
\gM_{\MM \NN} = (U_C)_{\MM}{}^{\KK}\GG_{\KK \LL} (U_C)_{\NN}{}^{\LL} 
\ee
where\footnote{I have regularised my generalised geometric conventions to maintain consistency between $\Gfour$ and $\Gsix$ expressions.
Accordingly I am using slightly different $\Gsix$ conventions to those in my previous paper \cite{Blair:2022gsx}.}
\be
\begin{split}
\GG_{\MM \NN} &= |\det \phi_{\ii\jj}|^{+1/3} \begin{pmatrix} 
\phi_{\ii\jj} & 0 & 0 \\ 
0 & 2 \phi^{\ii_1[\jj_1} \phi^{\jj_2]\ii_2} & 0 \\
0 & 0 & (\det \phi_{\ii \jj})^{-1} \phi_{\ii \jj} 
\end{pmatrix} \,,\\
(U_C)_{\MM}{}^{\NN}&= \begin{pmatrix}
\delta_\ii^\jj &  C_{\ii \jj_1\jj_2} & + \delta_{\ii}^{\jj} C_6\rmp\tfrac{1}{4!} \epsilon^{\jj \kk_1 \dots \kk_5} C_{\ii\kk_1 \kk_2} C_{\kk_3\dots \kk_5}\\
0 & 2 \delta_{\jj_1}^{[\ii_1} \delta^{\ii_2]}_{\jj_2} & \rmp \tfrac{1}{3!} \epsilon^{\jj \ii_1 \ii_2 \kk_1 \dots \kk_3}C_{\kk_1\kk_2\kk_3}  \\
0  & 0  & \delta_{\ii}^{\jj} & \\
\end{pmatrix}\,.
\end{split} 
\label{UC6}
\ee
The dual polyvector variables are now a trivector, $\Omega^{\ii\jj\kk}$, as well as a six-vector, $\Omega^{\ii_1\dots \ii_6} \equiv \epsilon^{\ii_1 \dots \ii_6} \Omega$.
These can be introduced using the following $\Gsix$ valued matrix:
\be
(U_\Omega)_{\MM}{}^{\NN} 
 = 
\begin{pmatrix}
\delta_\ii^\jj &  0 & 0\\
\Omega^{\ii_1\ii_2 \jj} & 2 \delta_{\jj_1}^{[\ii_1} \delta^{\ii_2]}_{\jj_2} & 0 \\
\delta_\ii^\jj \Omega \rmp \tfrac{1}{4!} \epsilon_{\ii\kk_1\dots\kk_5} \Omega^{\jj \kk_1 \kk_2} \Omega^{\kk_3\kk_4\kk_5}  &  
\rmp \tfrac{1}{3!} \epsilon_{\ii \jj_1\jj_2 \kk_1 \kk_2\kk_3} \Omega^{\kk_1 \kk_2\kk_3} 
 & \delta_{\ii}^{\jj} & \\
\end{pmatrix}\,.
\label{UOmega6}
\ee
I will use matrices of this form to describe the action of six-vector deformations. I can also embed the trivector appearing in the $\Gfour$ generalised frame in a $\Gsix$ matrix of the form \eqref{UOmega6}.
Note that trivector and six-vector transformations commute.

The complete $\Gsix$-covariant description of 11-dimensional supergravity also requires fields carrying five-dimensional indices \cite{Hohm:2013vpa}.
For instance, the 11-dimensional metric is decomposed as:
\be
\dd s^2_{11} = |\phi|^{-1/3} g_{\mu\nu} \dd x^\mu \dd x^\nu + \phi_{\ii\jj} D x^{\ii} D x^{\jj} \,,
\label{sixmetricdecomp}
\ee
where $D x^{\ii} \equiv \dd x^{\ii} + A_{(1)}^{\ii}$. The five-dimensional metric $g_{\mu\nu}$ is invariant under $\Gsix$. 
In addition there are gauge fields. The most relevant is the one-form $\Aa_{(1)}^{\MM}$ which contains the `Kaluza-Klein vector' $A_{(1)}^{\ii}$ appearing in the metric decomposition, as well as the components of the three- and six-form carrying one five-dimensional index. (The subscript $(p)$ now denotes a $p$-form on $\text{M}_5$, the five-dimensional part of the spacetime.)
I will only need the two-form field strength of this gauge field\footnote{The next field strength, $\mathcal{H}_{(3)\MM}$, is related to $\mathcal{F}_{(2)}{}^{\MM}$ by a notion of Hodge duality, hence does not contain additional physical information.}, which on general grounds can be written as:
\be
\mathcal{F}_{(2)}{}^{\MM} =
\begin{pmatrix}
F_{(2)}{}^{\ii} \\
 \Fdef_{(2) \ii_1 \ii_2} - C_{\ii_1 \ii_2 \jj} F_{(2)}{}^{\jj} \\
 \Fdef_{(2) \ii_1 \dots \ii_5} \rpm \tfrac{5!}{3!2!} C_{[\ii_1 \ii_2\ii_3} \Fdef_{|(2)| \ii_4\ii_5]} 
 + F_{(2)}{}^{\jj} ( C_{\ii_1 \dots \ii_5 \jj} \rmp 5 C_{\jj[\ii_1\ii_2} C_{\ii_3\ii_4\ii_5]})
\end{pmatrix} 
\label{sixF}
\ee
Here $F_{(2)}{}^\ii$ is a field strength for $A_{(1)}^{\ii}$, and $\Fdef_{(2) \ii\jj}$ and $\Fdef_{(2)\ii_1 \dots \ii_5}$ are related to components of the 11-dimensional four-form and dual seven-form ($F_{(7)} = \star F_{(4)}$) by field redefinitions involving $A_{(1)}^{\ii}$,  as in equation \eqref{Fdecomp_compact}. 
This general form of $\mathcal{F}_{(2)}{}^{\MM}$ is determined by the requirement that it transform covariantly under generalised diffeomorphisms, which fixes it to be `twisted' by the potentials in the same manner as the generalised metric.

\section{11-dimensional solutions and six-vector deformations from TsT}

\subsection{11-dimensional solutions from $\text{M}_7 \times \text{S}^3$} 
\label{generalmap}

I now write down the generalised U-duality solution generating mechanism \eqref{genUmap}. 
The existence of this map was pointed out in \cite{Blair:2020ndg}, and demonstrated for a particular example in \cite{Blair:2022gsx}, but a general expression for the resulting 11-dimensional backgrounds was not given. 
In appendix \ref{appredup} I derive the map for general pure NSNS solutions of type IIA SUGRA on $\text{S}^3$ that admit a consistent truncation to $\mathrm{CSO}(4,0,1)$ gauged supergravity.
In this section, I restrict to the case of a direct product solution $\text{M}_7 \times \text{S}^3$, with string frame metric and three-form field strength given by\footnote{For convenience, I have set the radius of the sphere to 1. To restore it in the 11-dimensional solution, make the rescaling $x^i \rightarrow x^i/R$ in all the components of the 11-dimensional fields.}
\be
\dd s^2_{s} = G_{\hI \hJ} \dd x^{\hI} \dd x^{\hJ} + \dd s^2_{\text{S}^3} \,,\quad
H_{(3)} = \Fthree_{(3)} +2 \mathrm{Vol}_{\text{S}^3} \,,
\quad
\Fthree_{(3)} \equiv \tfrac{1}{3!}
\Fthree_{\hI \hJ \hK} \dd x^{\hI}\wedge \dd x^{\hJ}\wedge \dd x^{\hK} \,,
\label{origbackIIA}
\ee
with $\hI,\hJ=0,\dots,6$.
In \eqref{origbackIIA}, I suppose that the seven-dimensional metric, $G_{\hI\hJ}$, the seven-dimensional restriction of the three-form, $\Fthree_{(3)}$, and the dilaton, $\varphi$, are independent of the coordinates on the sphere. 
This background can then be consistently truncated to a solution of seven-dimensional $\mathrm{CSO}(4,0,1)$ gauged maximal supergravity, following \cite{Cvetic:2000ah}, and then uplifted to a new solution of 11-dimensional supergravity, following \cite{Blair:2020ndg, Blair:2022gsx}.
This uplift uses the EDA generalised frame \eqref{fiveEDAframe} with data \eqref{moredata}.
The process of uplift in $\Gfour$ covariant variables followed by extraction of 11-dimensional fields is recounted in appendix \ref{appredup}.

The resulting 11-dimensional metric has the form:
\be
\begin{split}
\dd s_{11}^2 & = \left( e^{-2  \varphi} + \rho^2 \right)^{1/3} e^{-2\varphi/3}  G_{\hI \hJ} \dd x^{\hI} \dd x^{\hJ} 
 + \frac{\left(  \delta_{ij} + e^{2 \varphi} x_i x_j \right)}{\left( e^{-2 \varphi} + \rho^2 \right)^{2/3} } e^{-2\varphi/3}  \dd x^i \dd x^j\,,
\end{split}
\label{grand11metric}
\ee 
and the field strength and its dual are:
\be
\begin{split} 
F_{(4)} & = \Fthree_{(3)} \wedge x_i \dd x^i \rmp \tFfour_{(4)} 
+ \dd C_{\text{int}}\,,
\\
\star F_{(4)}  & = 
e^{-2\varphi}(4 e^{-2\varphi} +2 \rho^2) \mathrm{Vol}_7(G)
+ (\star_7 \dd \ln e^{-2\varphi}) \wedge x_i \dd x^i
\\  & \qquad
+ C_{\text{int}} \wedge \tFfour_{(4)} 
\rpm  \frac{e^{-2\varphi}}{e^{-2\varphi} + \rho^2}\Fthree_{(3)}\wedge \tfrac{1}{4!} \epsilon_{i_1 \dots i_4} \dd x^{i_1}\wedge \dots \wedge \dd x^{i_4} \,,
\end{split}
\label{grand11Fs}
\ee
where $\rho^2 \equiv \delta_{ij} x^i x^j$, and
\be
C_{\text{int}} =  \frac{\tfrac{1}{3!} \epsilon_{lijk} x^l}{e^{-2\varphi}+\rho^2} \dd x^i \wedge \dd x^j \wedge \dd x^k \,,
\ee
\be
\tFfour_{(4)} \equiv  
\tfrac{1}{3!} e^{-2\varphi} \sqrt{| G|} \epsilon_{\hI_1 \dots \hI_4 \hJ_1 \dots \hJ_3}  G^{\hJ_1 \hK_1}  G^{\hJ_2 \hK_2}  G^{\hJ_3 \hK_3} \Fthree_{\hK_1 \hK_2 \hK_3} \dd x^{\hI_1} \wedge\dots\wedge \dd x^{\hI_4}\,,
\ee
\be
(\star_7 \dd \ln e^{-2\varphi}) \equiv  \tfrac{1}{6!} e^{-2\varphi} \sqrt{|G|} \epsilon_{\hI_1 \dots \hI_6 \hJ} G^{\hJ \hK} \partial_{\hK} \ln e^{-2\varphi}  \dd x^{\hI_1} \wedge\dots\wedge \dd x^{\hI_6} \,,
\ee
\be
\mathrm{Vol}_7(G) \equiv \tfrac{1}{7!} \sqrt{|G|}  \epsilon_{\hI_1\dots \hI_7} \dd x^{\hI_1} \wedge\dots\wedge \dd x^{\hI_7}\,.
\ee
The Bianchi identity for $F_{(4)}$ implies it is possible to introduce a two-form $B_{(2)}$ and a three-form $\widetilde{B}_{(3)}$ such that
\be
\dd B_{(2)} = \Fthree_{(3)} \,,\quad
\dd \widetilde B_{(3)} = \tFfour_{(4)} \,.
\ee
In the initial type IIA picture, the two-form $B_{(2)}$ is the original NSNS two-form gauge field (restricted to $M_7$), while $\widetilde B_{(3)}$ arises from the dual six-form (wrapping the $\text{S}^3$).
Using these, the 11-dimensional three-form can be expressed as:
\be
C_{(3)} = B_{(2)} \wedge x_i \dd x^i 
\rmp \widetilde B_{(3)} + C_{\text{int}}\,.
\label{grandC}
\ee
In applications below, I will need to make explicit use of the dual six-form. 
This is defined by $\dd C_{(6)} = \star F_{(4)}  \rpm \tfrac12 C_{(3)} \wedge F_{(4)}$.
A direct calculation produces:
\be
\begin{split}
\dd C_{(6)} & = 
4 e^{-4\varphi}   \mathrm{Vol}_7(G)
\rpm \tfrac12 \widetilde B_{(3)} \wedge \tFfour_{(4)} 
\\ & \quad + \tfrac{1}{2} \dd \left( \rho^2 [  (\star_7 \dd \ln e^{-2\varphi} )  - \tfrac12 B_{(2)} \wedge \tFfour_{(4)} + \tfrac12 \Fthree_{(3)} \wedge \widetilde B_{(3)} ]\right)
\\ &  \quad + \tfrac{1}{2} \rho^2 \left(- \dd  (\star_7 \dd \ln e^{-2\varphi} )  +\Fthree_{(3)} \wedge \tFfour_{(4)}  + 4 e^{-2\varphi} \mathrm{Vol}_7(G)\right)
\\ &\quad + \dd  \left( \tfrac12 \widetilde B_{(3)} \wedge C_{\text{int}} \rpm \tfrac12 B_{(2)} \left( \frac{e^{-2\varphi}}{e^{-2\varphi}+\rho^2}  + 1 \right) \right) \wedge \tfrac{1}{4!} \epsilon_{i_1 \dots i_4} \dd x^{i_1}\wedge \dots \wedge \dd x^{i_4} \,.
\end{split}
\label{dC6}
\ee
The third line vanishes by the original type IIA equation of motion for the dilaton $\varphi$.
The first line and second line together define the components of $C_{(6)}$ restricted to the 7-dimensional part of spacetime inherited from the type IIA geometry.
The final line defines the mixed components of $C_{(6)}$ which include both 7-dimensional and 4-dimensional contributions.
The crucial part which will appear often below is the `internal' contribution proportional to the volume form on the 4-dimensional space:
\be
C_{(6) \text{int}} = \rpm \tfrac12 \left( \frac{e^{-2\varphi}}{e^{-2\varphi}+\rho^2} + 1 \right)B_{(2)}\wedge\tfrac{1}{4!} \epsilon_{i_1 \dots i_4} \dd x^{i_1}\wedge \dots \wedge \dd x^{i_4} \,.
\label{C6int}
\ee
It can be convenient in calculations to define spherical coordinates, letting $x^i = \rho \mu^i$ with $\delta_{ij} \mu^i \mu^j = 1$.
The following elementary results:
\be
\tfrac{1}{3!} \epsilon_{ijkl} x^i \dd x^j \wedge \dd x^k \wedge \dd x^l 
= \rho^4 \mathrm{Vol}_{\text{S}^3} \,,\quad
\tfrac{1}{4!} \epsilon_{ijkl}\dd x^i \wedge \dd x^j \wedge \dd x^k \wedge \dd x^l 
= \rho^3 \dd \rho \wedge \mathrm{Vol}_{\text{S}^3} \,,
\ee
along with $x_i \dd x^i = \rho \dd \rho$, allow the solution to be rewritten in these coordinates.
The metric and field strength are then:
\be
\begin{split}
\dd s_{11}^2 & = \left( e^{-2  \varphi} + \rho^2 \right)^{1/3} \left( e^{-2\varphi/3}  G_{\hI \hJ} d x^{\hI} d x^{\hJ} + e^{4\varphi/3} \dd \rho^2  \right)
 + \frac{e^{-2\varphi/3}\rho^2 \dd s^2_{\text{S}^3} }{\left( e^{-2 \varphi} + \rho^2 \right)^{2/3} }  \,,
 \\
F_{(4)} & = \Fthree_{(3)} \wedge \rho d \rho \rmp \tFfour_{(4)} + \dd \left( \frac{\rho^4}{e^{-2\varphi} + \rho^2} \mathrm{Vol}_{\text{S}^3} \right) \,.
\end{split}
\ee
A final comment is that the calculation of $d C_{(6)}$ also provides the Page charge density, $\mathcal{Q}_{\text{Page}} = \star F_{(4)} \rpm \tfrac{1}{2} C_{(3)} \wedge F_{(4)}$.
Integrating this over (spatial) seven-cycles calculates the M2 charge.
From \eqref{dC6} it follows that the Page charge is a total derivative and so vanishes up to large gauge transformations.
On the other hand, the form of $F_{(4)}$ suggests the presence of M5 charge.

\subsection{TsT to six-vector deformations: AdS${}_3 \times \text{T}^4 \times \text{S}^3$ examples}
\label{maptst}

My main goal in this paper is to study how T-duality transformations of the original type IIA solution then generate additional 11-dimensional solutions.
I will restrict to T-duality transformations which preserve the form of the ansatz \eqref{origbackIIA}: this means that I will only consider transformations which do not touch the three-sphere.
These transformations will be elements of $\Odd\subseteq \mathrm{O}(7,7)$ acting non-trivially on the metric $G_{\hI \hJ}$, 2-form $B_{\hI \hJ}$ and dilaton $\varphi$.

The observation underlying this goal was the discovery in \cite{Blair:2022gsx} that an $\mathrm{O}(2,2)$ TsT transformation of AdS${}_3 \times$ T$^4 \times$ S$^3$ was transmuted into a more complicated six-vector transformation of the dual 11-dimensional solution.
This six-vector deformation could be realised as an $\Gsix$ element. The unusual feature about this $\Gsix$ transformation was that it acted both on isometric and non-isometric directions.
A natural conjecture is that this would be true for other TsT transformations. 

In the next subsection, I will present a general demonstration of this fact applicable to any background of the form \eqref{origbackIIA} in which $\text{M}_7$ admits two commuting isometries. 
However, here I first focus on this AdS${}_3$ case, in order to motivate this goal and review the observation of \cite{Blair:2022gsx}. 
The original type IIA background is:
\be
\begin{split}
\dd s^2 &= \dd s^2_{\text{AdS}_3} + \dd s^2_{\text{T}^4} +
\dd s_{\text{S}^3} 
 \,,\quad B  = \tfrac12 r^2 \epsilon_{\alpha \beta} \dd x^\alpha \wedge \dd x^\beta + 2 b\,,\quad
e^{-2\varphi} = 1 \,,
\label{AdSTS}
\end{split}
\ee
with $\dd b = \mathrm{Vol}_{\text{S}^3}$ and where I take the AdS metric to be
\be
\dd s^2_{\text{AdS}_3}=
r^2 \eta_{\alpha\beta} \dd x^\alpha \dd x^\beta + \frac{\dd r^2}{r^2} \,,
\ee
with $\eta_{\alpha\beta}$ the 2-dimensional Minkowski metric.
This evidently is of the form \eqref{origbackIIA} with
\be
G_{\hI \hJ} \dd x^{\hI} \dd x^{\hJ} = 
\dd s^2_{\text{AdS}_3} 
 + \dd s^2_{\text{T}^4}  \,,\quad
 \Fthree_{(3)} = 2 \mathrm{Vol}_{\text{AdS}} \,,
\ee
and so gives an 11-dimensional solution (fitting into the class of solutions found in \cite{Lozano:2020bxo}) with
\be
\begin{split}
\dd s_{11}^2 & = \left(1+ \rho^2 \right)^{1/3}  (\dd s^2_{\text{AdS}_3} + \dd s^2_{\text{T}^4} )
 + \frac{\left(  \delta_{ij} +  x_i x_j \right)}{\left( 1 + \rho^2 \right)^{2/3} }  \dd x^i \dd x^j\,,
\end{split}
\ee 
\be
\begin{split}
F_{(4)} & = 2 \mathrm{Vol}_{\text{AdS}} \rpm 2 \mathrm{Vol}_{\text{T}^4} 
+ \dd \left(  \frac{\tfrac{1}{3!} \epsilon_{lijk} x^l}{1+\rho^2} \dd x^i \wedge \dd x^j \wedge \dd x^k \right)\,.
\end{split}
\ee
I now will deform the background \eqref{AdSTS} by carrying out three types of TsT deformations, using the formulae of \eqref{bivrules} and carry the result over to 11-dimensional solutions.

\subsubsection*{AdS-AdS deformation}

This is the deformation already analysed in \cite{Blair:2022gsx}.
The $\Odd$ bivector has only non-zero component $\beta^{\alpha \beta} =-\tfrac12 \lambda \epsilon^{\alpha \beta}$.
In 10 dimensions this produces: 
\be
\begin{split}
\dd s^2 &= \frac{r^2}{1+\lambda r^2} \eta_{\alpha\beta} \dd x^\alpha \dd x^\beta + \frac{\dd r^2}{r^2} + \dd s^2_{\text{T}^4} +
\dd s_{\text{S}^3} 
 \,,\\
B & = \tfrac12 \frac{r^2}{1+\lambda r^2} \epsilon_{\alpha \beta} \dd x^\alpha \wedge \dd x^\beta + 2 b\,,\quad
e^{-2\varphi} = 1+\lambda r^2 \,,
\label{AdSTT}
\end{split}
\ee
which (after setting $\lambda=1$ by a coordinate rescaling) corresponds to the extremal F1-NS5 brane solution in the near horizon limit of the NS5.
In eleven dimensions the result is:
\be
\begin{split} 
\dd s_{11}^2  &=
 (1+\rho^2 + \lambda r^2)^{1/3} (1+\lambda r^2)^{-2/3} r^2 \eta_{\alpha\beta} \dd x^\alpha \dd x^\beta
\\&\qquad +  (1+\rho^2 + \lambda r^2)^{1/3} (1+\lambda r^2)^{1/3} \left( \frac{\dd r^2}{r^2} + \dd s^2_{\text{T}^4} \right)
 \\& \qquad +   (1+\rho^2 + \lambda r^2)^{-2/3} (1+\lambda r^2)^{1/3} \left( \delta_{ij} + \frac{x_ix_j}{1+\lambda r^2} \right) \dd x^i \dd x^j \,,\\
 F_{(4)} & = \frac{2r x_i}{(1+\lambda r^2)^2} \tfrac{1}{2} \epsilon_{\alpha \beta} \dd x^\alpha \wedge \dd x^\beta \wedge \dd r \wedge \dd x^i \rpm 2 \mathrm{Vol}_{\text{T}^4} 
 + \dd \left(\frac{\epsilon_{lijk} x^l \dd x^i \wedge \dd x^j \wedge \dd x^k}{1+\lambda r^2 + \rho^2} \right)\,.
\end{split} 
\label{solutionDeformationX}
\ee
It was shown in \cite{Blair:2022gsx} how to view this as a six-vector deformation of the $\lambda = 0$ solution, with 
\be
\Omega^{\alpha \beta ijkl} = \tfrac12 \lambda \epsilon^{\alpha \beta ijkl} \,.
\ee
Verifying this involves making a 5+6 dimensional split of the coordinates, with $x^{\ii} = (x^\alpha, x^i)$ corresponding to the directions on which $\Gsix$ acts.
Under this decomposition, the five-dimensional metric and the generalised metric suffice to consider all physical degrees of freedom, as the two- and three-form field strengths are zero and the four-form field strength is dual (in five dimensions) to the derivative of the generalised metric.
The $\lambda$-dependence on the generalised metric can then be shown to correspond to a six-vector deformation (in \cite{Blair:2022gsx} this was done by first switching to spherical coordinates $x^i \rightarrow (\rho, \theta^\alpha)$, and noting there was a natural 3+3 decomposition of the background in terms of $(x^\alpha,\rho)$ and $(\theta^\alpha)$. For such backgrounds, general formulae for the action of the six-vector deformation on the generalised metric can be straightforwardly obtained and applied -- see the appendix of \cite{Blair:2022gsx}).

\subsubsection*{AdS-torus deformation}

I now describe a deformation mixing the AdS${}_3$ and torus part of the geometry. 
I write the time and spatial isometry directions within the AdS${}_3$ as $t$ and $z$, and let $y$ denote one of the torus directions.
I then take the $\Odd$ bivector to have only non-zero component $\beta^{z y} =\gamma$.
In 10 dimensions, this produces
\be
\begin{split}
\dd s^2 &= \frac{r^2}{1+\gamma^2 r^2}(- (\dd t + \gamma \dd y)^2 + \dd z^2 ) + \frac{\dd r^2}{r^2} + \dd s^2_{\text{T}^4} +
\dd s_{\text{S}^3} 
 \,,\\
B =&  \frac{r^2}{1+\gamma^2 r^2}( \dd t +\gamma \dd y) \wedge \dd z+ 2 b\,,\quad
e^{-2\varphi} = 1+\gamma^2 r^2 \,.
\label{AdSJT}
\end{split}
\ee
The new 11-dimensional metric is
\be
\begin{split}
\dd s_{11}^2 & = \left( 1 + \gamma^2 r^2 + \rho^2 \right)^{1/3} (1 + \gamma^2 r^2)^{-2/3}
\left(
 - r^2 ( \dd t + \gamma \dd y )^2 + r^2 \dd z^2 
 \right) 
\\ & + 
\left( 1 + \gamma^2 r^2 + \rho^2 \right)^{1/3} (1 + \gamma^2 r^2)^{1/3}
\left(
 \frac{\dd r^2}{r^2} 
+ \dd s^2_{\text{T}^4}
\right) 
\\  & 
 + \left( 1 + \gamma^2 r^2+ \rho^2 \right)^{-2/3} (1 + \gamma^2 r^2)^{1/3}  {\left(  \delta_{ij} + \frac{x_i x_j}{1 + \gamma^2 r^2} \right)} \ \dd x^i \dd x^j
 \,,
\end{split}
\ee 
and the four-form is
\be
\begin{split}
F_{(4)} & = \frac{2r}{(1+\gamma^2 r^2)^2} (\dd t+\gamma \dd y) \wedge \dd z \wedge \dd r \wedge x_i \dd x^i
\rpm 2 \mathrm{Vol}_{\text{T}^4} 
\\ & \qquad
 + \dd \left(  \frac{\tfrac{1}{3!} \epsilon_{lijk} x^l}{1+\gamma^2 r^2 +\rho^2} \dd x^i \wedge \dd x^j \wedge \dd x^k \right)\,.
\end{split}
\ee
The coordinate transformation $t' = t+\gamma y$ shows that this solution is identical to \eqref{solutionDeformationX} with the identification $\lambda = \gamma^2$.
Similarly the 10-dimensional background \eqref{AdSJT} is identical to \eqref{AdSTT} on the same coordinate transformation.
The fact that these apparently distinct duality transformations of the original AdS background lead to the same deformed backgrounds was commented on in \cite{Chakraborty:2020xyz}.
This could also be viewed as a self-duality of AdS${}_3\times$ T$^4 \times \text{S}^3$ under a particular sequence of transformations of the form TsTsTsT i.e. TsT/coordinate shift/TsT.
I discuss this in more detail in appendix \ref{genkill}.

\subsubsection*{Torus-torus deformation}

I finally take the $\Odd$ bivector to have only non-zero components in the torus directions.
I label these as $(y^1,y^2,y^3,y^4)$, and consider the case $\beta^{y^1y^2} =\lambda$.
In 10 dimensions, this produces
\be
\begin{split}
\dd s^2 &=\dd s^2_{\text{AdS}_3}  + \frac{1}{1+\lambda^2} \dd s^2_{\text{T}^2}(12)  + \dd s^2_{\text{T}^2}(34)+ 
\dd s_{\text{S}^3} 
 \,,\\
B =&  \frac{1}{2} r^2\epsilon_{\alpha \beta} \dd x^\alpha \wedge \dd x^\beta - \frac{\lambda}{1+\lambda^2} \dd y^1 \wedge \dd y^2 + 2 b\,,\quad
e^{-2\varphi} = 1+\lambda^2 \,.
\label{AdSJJ}
\end{split}
\ee
The new 11-dimensional metric is
\be
\begin{split}
\dd s_{11}^2 & = 
\left( 1+\lambda^2 + \rho^2 \right)^{1/3} (1+\lambda^2)^{-2/3} 
 \dd s^2_{T^2}(12)
 \\ & \qquad + \left( 1+\lambda^2 + \rho^2 \right)^{1/3} (1+\lambda^2)^{1/3} \left(
\dd s^2_{\text{AdS}_3} + \dd s^2_{T^2}(34) 
\right)
\\ & \qquad  + \left( 1+\lambda^2 + \rho^2 \right)^{-2/3}(1+\lambda^2)^{1/3} \left(  \delta_{ij} + \frac{x_i x_j}{1+\lambda^2} \right)  \dd x^i \dd x^j\,,
\end{split}
\label{torustorus}
\ee 
and the field strength is
\be
\begin{split}
F_{(4)} & =  2 \mathrm{Vol}_{\text{AdS}_3}  \wedge x_i\dd x^i
\rpm 2 \mathrm{Vol}_{\text{T}^4} 
+ \dd \left(  \frac{\tfrac{1}{3!} \epsilon_{lijk} x^l}{1+\lambda^2+\rho^2} \dd x^i \wedge \dd x^j \wedge \dd x^k \right)\,.
\end{split}
\ee
This can again be viewed as a six-vector deformation, but is a more complicated case than the AdS-AdS deformation.
Making again a 5+6 dimensional split, the directions on which $\Gsix$ acts are $x^{\ii} = (y^1,y^2, x^i)$.
As the field strength $F_{(4)}$ and its dual have components proportional to $\mathrm{Vol}_{\text{T}^4}$, this decomposition leads to a non-vanishing two-form field strength of the form \eqref{sixF}.
It can then be shown, using similar techniques to those applied for the AdS-AdS deformation in \cite{Blair:2022gsx} that the five-dimensional metric, generalised metric and this field strength can all be viewed as a six-vector deformation of the $\lambda = 0$ solution (these fields suffice to capture all physical degrees of freedom), where the six-vector has components:
\be
\Omega^{y^1 y^2 ijkl} = - \lambda \epsilon^{ijkl} \,.
\ee
Rather than describe the particulars of this example in detail, I instead move on to the general explanation of how these six-vector deformations are induced. 

\subsection{TsT to six-vector deformations: general case}
\label{tstgeneral}

I now show how to understand the action of $\mathrm{O}(2,2)$ transformations on the 11-dimensional background given by \eqref{grand11metric} and \eqref{grand11Fs}.
I suppose that the 7-dimensional metric $G_{\hI \hJ}$ admits two abelian isometry directions, with adapted coordinates $x^\alpha$.
I again package these coordinates with the four-dimensional coordinates $x^i$ to produce the six-dimensional coordinates $x^{\ii}=(x^\alpha, x^i)$.
I denote the remaining five coordinates by $x^\mu$.
After making a Kaluza-Klein decomposition of the metric $G_{\hI \hJ}$, of the form \eqref{10metricdec}, the line element \eqref{grand11metric} becomes
\be
\begin{split} 
\dd s^2_{11}& = (e^{-2\varphi}+\rho^2)^{1/3} e^{-2\varphi/3} G_{\mu\nu}  \dd x^\mu \dd x^\nu 
+ 
(e^{-2\varphi}+\rho^2)^{1/3} e^{-2\varphi/3}  G_{\alpha \beta} D x^\alpha D x^\beta
\\ & \qquad
+ \frac{\left(  \delta_{ij} + e^{2 \varphi} x_i x_j \right)}{\left( e^{-2 \varphi} + \rho^2 \right)^{2/3} } e^{-2\varphi/3}  \dd x^i \dd x^j \,,
\end{split}
\ee
where $D x^\alpha = \dd x^\alpha + A_\mu{}^\alpha \dd x^\mu$.
Using the general expression \eqref{sixmetricdecomp}, the $\Gsix$ invariant metric is then 
\be
g_{\mu\nu} = e^{-4 \gendil /3} G_{\mu\nu} \,,
\ee
where
\be
e^{-2\gendil} \equiv e^{-2\varphi} \sqrt{|\det G_{\alpha \beta}|} \,,
\ee
is the $\mathrm{O}(2,2)$ invariant dilation.
As $G_{\mu\nu}$ is also invariant under $\mathrm{O}(2,2)$, it follows that $g_{\mu\nu}$ is likewise invariant.

I now focus on the $\Gsix$ generalised metric encoding the six-dimensional components of the metric, three-form and six-form. 
Explicitly, these ingredients are:
\be
\phi_{\ii\jj} 
=\begin{pmatrix}
(e^{-2\varphi}+\rho^2)^{1/3} e^{-2\varphi/3} G_{\alpha \beta} & 0 \\
0 & (e^{-2\varphi}+\rho^2)^{-2/3} e^{-2\varphi/3} (\delta_{ij} + e^{2\varphi} x_ix_j ) 
\end{pmatrix}
\,,
\label{ingred1}
\ee
\be
C_{\alpha \beta i} = B_{\alpha \beta} x_i \,,\quad
C_{ijk} = \frac{\epsilon_{lijk} x^l}{e^{-2\varphi} + \rho^2}  \,,\quad
C_6 = \rpm  \tfrac12 B \left( \frac{e^{-2\varphi}}{e^{-2\varphi}+\rho^2} + 1 \right) \,,
\label{ingred2}
\ee
where I let $B_{\alpha\beta} \equiv B \epsilon_{\alpha\beta}$ as before.
The generalised metric components can then be straightforwardly, if tediously, computed. 
In order to give the results in a usable fashion, it helps to remember that the dependence on the $x^i$ coordinates must enter through a twist by a trivector, according to the $\Gfour$ EDA generalised frame \eqref{fiveEDAframe}, \eqref{moredata}.
This is easily embedded in $\Gsix$.
Define a trivector $\pi^{\ii\jj\kk}$ with $\pi^{ijk} = \epsilon^{ijkl} x_l$, $\pi^{ij\alpha} = \pi^{i\alpha\beta}=0$.
Then the $\Gsix$ generalised metric built using \eqref{ingred1} and \eqref{ingred2} can be verified to factorise as:
\be
\gM_{\MM \NN} (x^\mu,x^i)
= \UEDA_{\MM}{}^{\AA} (x^i) M_{\AA \BB} (x^\mu) \UEDA_{\NN}{}^{\BB}(x^i) 
\label{sixEDAfactor}
\ee
with\footnote{This identifies $\Omega^{\ii\jj\kk} = - \pi^{\ii\jj\kk}$ in \eqref{UOmega6}. This is because $\pi^{ijk}$ was originally defined in \eqref{fiveEDAframe} appearing in the $\Gfour$ generalised frame $\UEDA_A{}^M$ which is the inverse of $\UEDA_M{}^A$.}
\be
\UEDA_{\MM}{}^{\AA} (x^i) 
 = \begin{pmatrix}
 \delta_{\ii}{}^{\jj} &0 & 0  \\
- \pi^{\ii\ii' \jj} & 2 \delta^{\ii\ii'}_{\jj\jj'} & 0\\
0 & \rpm \tfrac{1}{3!} \epsilon_{\ii \jj\jj' \kk_1 \kk_2 \kk_3} \pi^{\kk_1 \kk_2 \kk_3}  & \delta_\ii{}^\jj
\end{pmatrix}\,. 
\label{UOmegaEDA}
\ee
I simplify expressions by avoiding the (at this point unnecessary) distinction between the components of the $\MM$ and $\AA$ indices. 

I then look at the generalised metric $M_{\AA \BB}$, describing the background of \eqref{grand11metric} with $x^i =0$. (This is not a solution on its own.) 
This can in turn be factorised as
\be
M_{\AA \BB} = (U_C)_{\AA}{}^{\CC} \mathcal{G}_{\CC \DD}  (U_C)_{\BB}{}^{\DD}
\ee 
where ${\mathcal{G}}$ and ${U}_{C}$ are defined as in \eqref{UC6} with 
\be
\phi_{\ii \jj} = \begin{pmatrix} e^{-4\varphi/3} G_{\alpha \beta} & 0 \\
0 & e^{2\varphi/3} \delta_{ij}
\end{pmatrix}\,,\quad
C_{\ii\jj\kk} = 0 \,,\quad
C_6 = B \,.
\ee
This generalised metric has the following block structure:
\be
M_{\AA \BB} = \begin{pmatrix}
M_{\ii \jj} & 0 & M_{\ii \bar \jj} \\
0 & M^{\ii\ii',\,\jj\jj'} & 0 \\
M_{\bar\ii \jj} & 0 & M_{\bar \ii \,	\bar \jj} 
\end{pmatrix} \,,
\ee
where, with $G \equiv \det G_{\alpha \beta}$,
\be
M_{\ii \jj} = 
\begin{pmatrix}
e^{-4 \gendil/3} G_{\alpha \beta} (1 + G^{-1} B^2 ) & 0 \\
0 & e^{2\gendil/3} \delta_{ij} |G|^{1/2} (1+ G^{-1} B^2) 
\end{pmatrix} \,,
\ee
\be
M_{\ii \bar \jj} = 
\begin{pmatrix}
\rpm e^{-4 \gendil/3} B G^{-1} G_{\alpha \beta}  & 0 \\
0 & \rpm e^{2\gendil/3} \delta_{ij} |G|^{1/2} B
\end{pmatrix} \,,
\ee
\be
M_{\bar\ii\, \bar \jj} = 
\begin{pmatrix}
e^{-4 \gendil/3} G^{-1} G_{\alpha \beta} & 0 \\
0 & e^{2\gendil/3} \delta_{ij} |G|^{1/2} 
\end{pmatrix} \,,
\ee
\be
M^{\alpha\beta,\gamma\delta} = \tfrac{G}{|G|} e^{8 \gendil/3} \epsilon^{\alpha \beta} \epsilon^{\gamma \delta}\,,\quad
M^{ij,kl} = 2 e^{-4\gendil/3} \delta^{i[k} \delta^{l]j}\,,\quad
M^{i\alpha,j\beta} =  e^{2\gendil/3}|G|^{1/2} G^{\alpha \beta} \delta^{ij} \,.
\ee
The structure of the generalised metric becomes clear on identifying the following tensor product structure: 
\be
\begin{pmatrix}
M_{\ii \jj}  & M_{\ii \bar \jj} \\
M_{\bar\ii \jj}  & M_{\bar \ii \,	\bar \jj} 
\end{pmatrix} 
=\mathcal{H}_\rho\otimes
\begin{pmatrix}
e^{-4\gendil/3} |G|^{-1/2} G_{\alpha \beta} & 0 \\
0 & e^{2\gendil/3} \delta_{ij} 
\end{pmatrix}  \,,
\ee
where $\mathcal{H}_\rho$ as defined in \eqref{Htaurho} is one of the $\mathrm{SL}(2)$ generalised metrics appearing in the factorisation $\mathrm{O}(2,2) \sim \mathrm{SL}(2)_{\tau} \times \mathrm{SL}(2)_{\rho}$. 
The other $\mathrm{SL}(2)$ generalised metric is $\mathcal{H}_\tau = |G|^{-1/2} G_{\alpha \beta}$, which also naturally appears in the above expressions.
The $\mathrm{SL}(2)_\tau$ T-dualities are geometric in nature and do not act in an interesting manner.
I therefore consider the fate of generic $\mathrm{SL}(2)_\rho$ T-dualities.
In the above decomposition, these act on $\mathcal{H}_\rho$, and leave invariant $|G|^{-1/2} G_{\alpha \beta}$, the generalised dilaton, and hence the rest of the $\Gsix$ generalised metric $M_{AB}$.

A general $\mathrm{SL}(2)_\rho$ transformation acting on $\mathcal{H}_\rho$ embeds into an $\Gsix$ transformation of $M_{\AA\BB}$ of the form:
\be
\mathcal{O}_{\AA}{}^{\BB}
 = \begin{pmatrix}
 a \delta_{\ii}^{\jj} &0 & b \delta_\ii^\jj  \\
0 & 2 \delta^{\ii\ii'}_{\jj\jj'} & 0\\
c\delta_{\ii}^{\jj} & 0  & d \delta_\ii^\jj
\end{pmatrix}\,.
\label{Usixabcd}
\ee
Applying this to $M_{\AA\BB}$ and then contracting with $\UEDA_{\MM}{}^{\AA}$ as in \eqref{sixEDAfactor} always produces a valid 11-dimensional solution. 
The transformation acting on the full generalised metric $\gM_{\MM\NN}$ is $\mathcal{O}_{\MM}{}^{\NN} = \UEDA_{\MM}{}^{\AA} \mathcal{O}_{\AA}{}^{\BB} \UEDA_{\BB}{}^{\NN}$, which has the form:
\be
\mathcal{O}_{\MM}{}^{\NN} 
= \begin{pmatrix}
a \delta_{\ii}^{\jj} & -b \tilde \pi_{\ii\jj_1\jj_2} & b \delta_{\ii}^{\jj} \\
\pi^{\ii_1 \ii_2\jj} (1-a) & 2 \delta^{[\ii_1}_{\jj_1} \delta^{\ii_2]}_{\jj_2} & -b \pi^{\ii_1\ii_2 \jj} \\
c \delta_{\ii}^{\jj} & \tilde \pi_{\ii\jj_1\jj_2} (1-d) & d\delta^{\ii\jj} 
\end{pmatrix}
\label{Utransmogrified}
\ee
where $\tilde \pi_{\ii\jj\kk} = \rpm \tfrac{1}{3!} \epsilon_{\ii\jj\kk \ll_1 \dots \ll_3} \pi^{\ll_1 \ll_2\ll_3}$.

The $\mathrm{O}(2,2)$ bivector shift \eqref{Hrholambda} is the case $a=1,d=1,b=0,c=-\lambda$.
This is the special case where $\mathcal{O}_{\MM}{}^{\NN}$ in \eqref{Utransmogrified} is constant, and 
corresponds to a $\Gsix$-valued six-vector transformation of the form
\be
(\mathcal{O}_\lambda)_{\MM}{}^{\NN}
 = \begin{pmatrix}
 \delta_{\ii}^{\jj} &0 & 0  \\
0 & 2 \delta^{\ii\ii'}_{\jj\jj'} & 0\\
-\lambda \delta_{\ii}^{\jj} & 0  & \delta_\ii^\jj
\end{pmatrix}\,.
\label{Usix}
\ee
This is identical to the transformation $\mathcal{O}_{\AA}{}^{\BB}$ acting on $M_{\AA \BB}$: this is due to the fact that the six-vector transformation commutes with the trivector twist \eqref{UOmegaEDA}.

It is interesting to also consider some other examples of T-duality transformations, for which the transformation $\mathcal{O}_\MM{}^{\NN}$ involves the coordinate dependent trivector. 
These are also examples of solution generating `generalised U-dualities'.
For example, the case $a=1,d=1,c=0$, $b\neq 0$ corresponds to a B-field shift.
The matrix $\mathcal{O}_{\MM}{}^{\NN}$ is upper triangular and comparing to \eqref{UC6} can be seen to correspond to the following shifts of the three-form and six-form:
\be
C_{\alpha \beta i} \rightarrow C_{\alpha \beta i} \rpm b \epsilon_{\alpha \beta} x_i \,,\quad
C_6 \rightarrow C_6 + b\,.
\ee
The case $a=0=d$, $b=\pm 1$, $c=\mp 1$ corresponds to Buscher duality on two directions, and gives the non-trivial $\Gsix$ transformation:
\be
\mathcal{O}_{\MM}{}^{\NN} 
= \begin{pmatrix}
0 & \mp  \tilde \pi_{\ii\jj_1\jj_2} & \pm \delta_{\ii}^{\jj} \\
\pi^{\ii_1 \ii_2\jj}  & 2 \delta^{[\ii_1}_{\jj_1} \delta^{\ii_2]}_{\jj_2} & \mp  \pi^{\ii_1\ii_2 \jj} \\
\mp \delta_{\ii}^{\jj} & \tilde \pi_{\ii\jj_1\jj_2}  & 0
\end{pmatrix}
\ee
I could also consider the Buscher duality transformation on one direction, which swaps $\tau$ and $\rho$. In general, I would not expect this to be realisable as a $\Gsix$ transformation, as this is a transformation which exchanges type IIA and type IIB, and the solution generating map \eqref{genUmap} specifically starts with a type IIA solution.

I next look at the transformation properties of the field strengths. 
I make the standard redefinition \eqref{Fdecomp_compact} of the field strengths. 
The components of the redefined field strengths carrying two five-dimensional indices are:
\be
\begin{split}
\Fdef_{\mu\nu \alpha \beta} & = \rmp \tfrac{1}{3!} e^{-2\gendil} \sqrt{|G|} \epsilon_{\mu\nu\alpha\beta \sigma_1 \dots \sigma_3} G^{\sigma_1 \lambda_1} G^{\sigma_2 \lambda_2} G^{\sigma_3 \lambda_3} \hdef_{\lambda_1\lambda_2\lambda_3}\,,\quad
\Fdef_{\mu\nu i \alpha}  = - \hdef_{\mu\nu \alpha} x_i \,,\\
\Fdef_{\mu\nu \alpha \beta  ijk} & = \Fdef_{\mu\nu\alpha\beta} C_{ijk}  \,,\quad
\Fdef_{\mu\nu \alpha i_1 \dots i_4}  = \rpm \hdef_{\mu\nu \alpha} \epsilon_{i_1 \dots i_4} \frac{e^{-2\varphi}}{e^{-2\varphi}+\rho^2} \,,
\end{split} 
\ee
where $\hdef_{\mu\nu\rho}$ and $\hdef_{\mu\nu\alpha}$ are obtained from the Kaluza-Klein decomposition of $\Fthree_{\hI\hJ\hK}$ mimicking \eqref{TFs}.

Using the expression for the $\Gsix$-covariant two-form field strength given by \eqref{sixF}, I can verify that these components appear in the following (twisted) field strength:
\be
\mathcal{F}_{\mu\nu}{}^{\mathsf{M}} = \UEDA_{\AA}{}^{\MM} \mathcal{F}_{\mu\nu}{}^{\AA} \,,\quad
\mathcal{F}_{\mu\nu}{}^{\AA} 
= 
\begin{pmatrix}
F_{\mu\nu}{}^\alpha \\ 0 \\
\Fdef_{\mu\nu\alpha\beta} \\ 0 \\ 0 \\
\rpm\epsilon_{i_1 \dots i_4}(  \hdef_{\mu\nu\alpha}  + B_{\alpha \beta}F_{\mu\nu}{}^\beta  ) \\
0
\end{pmatrix} \,.
\ee
The effect of the trivector twist is simply to induce a non-zero component $\mathcal{F}_{\mu\nu \alpha i} = x_i (\hdef_{\mu\nu\alpha}+B_{\alpha\beta} F_{\mu\nu}{}^\beta)$.

Referring to \eqref{TF} I identify the $\mathrm{O}(2,2)$ doublet $(F_{\mu\nu}{}^\alpha, \hdef_{\mu\nu\alpha} +B_{\alpha\beta}F_{\mu\nu}{}^\beta)$, and note that $\hdef_{\mu\nu\rho}$, and hence $\Fdef_{\mu\nu\alpha\beta}$, is a singlet.
For $\mathrm{SL}(2)_\rho$ T-dualities, this implies that $\mathcal{F}_{\mu\nu}{}^{\AA}$ transforms via (the inverse of) \eqref{Usixabcd} as expected.
In particular, bivector transformations of the $\mathrm{O}(2,2)$ doublet induce exactly the expected six-vector transformations acting on $\mathcal{F}_{\mu\nu}{}^{\MM}$.

In summary, the calculations in this subsection demonstrate that bivector transformations of the original type IIA solution are mapped to the $\Gsix$ six-vector transformations of the form \eqref{Usix}. 
Unlike general $\mathrm{SL}(2)_\rho$ T-dualities, these commute with the trivector twist defining the 11-dimensional solution, and so can be viewed as constant $\Gsix$ transformations.
Nonetheless these transformations have an unusual character, in that they involve the would-be U-duality group $\Gsix$ acting in both isometric and non-isometric directions.

\section{Algebraic interpretation of the deformation} 
\label{algint} 

I now discuss in more detail the geometric and algebraic characterisation of these six-vector transformations.

\subsubsection*{The six-vector deformation and $\Gsix$ EDA}

The non-isometric directions involved in the deformation are those of the four-dimensional geometry $\Mfour$. This geometry has special characteristics resulting from the underlying algebraic structure.
The failure of the geometry to admit translational isometries can be precisely characterised as follows.
Denoting by $g_{\hI \hJ}$ the 7-dimensional block of the metric, and $g_{ij}$ the four-dimensional block, the vector fields $v_a = \delta_a^i \partial_i$, $a=1,\dots,4$, associated to translations in $x^i$ are not Killing but rather obey \cite{Sakatani:2020iad}
\be
\begin{split}
L_{v_a} g_{\hI \hJ} &= - \tfrac{2}{3} \tfrac{1}{3!} \tilde f^{bcd}{}_a C_{bcd} g_{\hI \hJ} \,,\\
L_{v_a} g_{ij}& = - \tfrac{2}{3} \tfrac{1}{3!} \tilde f^{bcd}{}_a C_{bcd} g_{ij} + \tilde f^{bcd}{}_a C_{bc(i} g_{j) d}  \,,\\
\end{split}
\ee
where $C_{abc} = v_a{}^i v_b{}^j v_c{}^k C_{ijk}$ and $\tilde f^{bcd}{}_a = \epsilon^{bcde}\delta_{ea}$.
The vector fields $v_a$ are associated with the Lie algebra which is selected as the physical subalgebra of the $\mathrm{CSO}(4,0,1)$ exceptional Drinfeld algebra.
This corresponds to the abelian translational subalgebra of the Lie algebra of $\mathrm{CSO}(4,0,1)$.

Given an 11-dimensional geometry admitting two commuting Killing vectors, $k_{\alpha}$, and the set of four vectors $v_a$ associated to the EDA structure, I group these together and let
\be
V_{\as} = ( k_{\alpha}, v_a )\,.
\ee
Then, the six-vector parameter appearing in the 11-dimensional solutions can be naturally identified with 
\be
\Omega^{(6)} 
= r^{\as_1 \dots \as_6} V_{\as_1} \dots V_{\as_6}{} \,,
\label{naturalSixVec}
\ee
with $r^{\as_1 \dots \as_6} = \lambda \epsilon^{\as_1 \dots \as_6}$.
This gives a covariant formula for the six-vector deformation parameter.
In adapted coordinates for the isometries, and in the original coordinates used to define the 11-dimensional solution, this six-vector will be a constant.
In other choices of coordinates, such as spherical coordinates on $\Mfour$ as used in the analysis of the AdS-AdS deformation in \cite{Blair:2022gsx} it will not be constant.

The six-vector transformation can also be viewed as acting on the underlying $\mathrm{CSO}(4,0,1)$ exceptional Drinfeld algebra, embedded into an $\Gsix$ EDA.
The $\Gsix$ EDA is constructed in \cite{Malek:2020hpo}.
I list the resulting brackets for the special case where the only non-zero structure constants appearing are those of a 3-algebra, $f^{\as_1 \dots \as_3}{}_{\bs}$, with the assumption that $f^{\as_1 \as_2 \bs}{}_{\bs} = 0$:
\be
\begin{split} 
[ T_{\as} , T_{\bs} ] & = 0 \,,\quad
[ T^{\as_1 \as_2} , T^{\bs_1 \bs_2} ] =  2 f^{\as_1 \as_2 [ \bs_1 }{}_{\cs} T^{\bs_2 ] \cs}\,,\\
[ T_{\as}, T^{\bs_1 \bs_2} ] & = -f^{\bs_1 \bs_2 \cs}{}_{\as}T_{\cs} \,,\quad
 [ T^{\bs_1 \bs_2}, T_{\as} ]  =  f^{\bs_1 \bs_2 \cs}{}_{\as} T_{\cs}\,,
\\
[ T_{\as}, T^{\bs_1 \dots \bs_5} ] & = -10 f^{[\bs_1 \bs_2 \bs_3}{}_{\as} T^{\bs_4 \bs_5]} \,,\\
[ T^{\bs_1 \dots \bs_5} ,T_{\as} ] & =  10 f^{[\bs_1 \bs_2 \bs_3}{}_{\as} T^{\bs_4 \bs_5]}
+20 f^{[\bs_1 \bs_2 \bs_3}_{\cs} \delta^{\bs_4}_{\as} T^{\bs_5] c} 
 \,,
\\
[ T_{\as_1 \as_2}, T^{\bs_1 \dots \bs_5} ] & = -5 f^{\as_1 \as_2 [ \bs_1}{}_{\cs} T^{\bs_2 \dots \bs_5] \cs} \,,\quad
[ T^{\bs_1 \dots \bs_5},T_{\as_1 \as_2} ]  = +10 f^{[ \bs_1 \bs_2\bs_3}{}_{\cs} T^{\bs_4 \bs_5] \as_1 \as_2} \,,\\
[ T^{\as_1 \dots \as_5}, T^{\bs_1 \dots \bs_5} ] & = 0\,.
\end{split}
\label{sixEDA}
\ee
These are again not antisymmetric in general.

To connect transformations of the algebra with transformations of the geometry, I adopt the following general viewpoint.
Given the $\Gsix$ factorisation of the generalised metric, as in equation \eqref{sixEDAfactor},
and the associated structure constants $F_{\AA \BB}{}^{\CC}$ generated by the generalised frame $\UEDA_{\AA}$, I can consider a solution generating transformation of the following form:
\be
M'_{\AA \BB}= \mathcal{O}_{\AA}{}^{\CC} 
\mathcal{O}_{\BB}{}^{\DD} M_{\CC \DD}\,,\quad
F'_{\AA \BB}{}^{\CC}
= \mathcal{O}_{\AA}{}^{\DD}\mathcal{O}_{\BB}{}^{\EE}
(\mathcal{O}^{-1})_{\FF}{}^{\CC} F_{\DD \EE}{}^{\FF} \,,
\label{sixsgt}
\ee
for $\mathcal{O}_{\AA}{}^{\CC}$ a constant $\Gsix$ transformation.
This in turn implies a transformation of the $\Gsix$ EDA algebra, with the generators transforming as $T_{\AA} \rightarrow T'_{\AA} = \mathcal{O}_{\AA}{}^{\BB} T_{\BB}$.
If I can find a generalised frame $\EE'_{\AA}$ realising a generalised parallelisation based on the $F'_{\AA \BB}{}^{\CC}$ structure constants, then I can construct a new 10 or 11-dimensional background with
$\gM'_{\MM\NN} = E'_{\MM}{}^{\AA} E'_{\NN}{}^{\BB} M'_{\AA \BB}$.

I can then consider the action of the general $\mathrm{SL}(2)_\rho$ transformation \eqref{Usixabcd}. 
This leaves invariant the algebra \eqref{sixEDA}.
To see this, it is illuminating to rewrite \eqref{sixEDA} on defining $T_{\bar \as} = \tfrac{1}{5!} \epsilon_{\as \bs_1 \dots \bs_5} T^{\bs_1 \dots \bs_5}$.
Letting also $\tilde f_{\as_1 \as_2 \as_3,\bs} \equiv \tfrac{1}{3!} \epsilon_{\as_1\as_2\as_3 \cs_1 \cs_2 \cs_3} \tilde f^{\cs_1\cs_2\cs_3}{}_{\bs}$, the algebra is:
\be
\begin{split}
[T_{\as}, T_{\bs}] & = 0 \,,\quad [T_{\as}, T^{\bs_1 \bs_2}]  = - \tilde f^{\bs_1 \bs_2 \cs}{}_{\as} T_{\cs} = - [T^{\bs_1 \bs_2},T_{\as}]  \,,\\ 
[T_{\bar\as}, T_{\bar\bs}] & = 0 \,,\quad [T_{\bar\as}, T^{\bs_1 \bs_2}]  = - \tilde f^{\bs_1 \bs_2 \cs}{}_{\as} T_{\bar\cs} = - [T^{\bs_1 \bs_2},T_{\bar\as}]  \,,\\ 
[T_\as, T_{\bar\bs}] & = \tfrac12 \tilde f_{\bs \cs \ds,\as} T^{\cs \ds} \,,\\
[T_{\bar\bs}, T_{\as}] & = -\tfrac12 \tilde f_{\bs \cs \ds,\as} T^{\cs \ds}- \tilde f_{\as\bs\cs,\ds} T^{\cs \ds} \,,\\
[T^{\as_1\as_2}, T^{\bs_1\bs_2} ] & = 2 \tilde f^{\as_1 \as_2 [\bs_1}{}_{\cs} T^{\bs_2]\cs} \,.
\end{split} 
\ee
The condition that $\tilde f^{\as \bs \cs}{}_{\cs} = 0$ translates to $\tilde f_{[\as\bs\cs,\ds]} =0$, which implies that $[T_{\as}, T_{\bar\bs}] = - [T_{\bar \as}, T_{\bs}]$.
Using this it can be checked that $(T_{\as}, T_{\bar \as})$ form an $\mathrm{SL}(2)$ doublet, and the algebra is invariant under $(T_{\as}, T_{\bar \as}) \rightarrow ( a T_{\as} + b T_{\bar \as}, d T_{\bar \as} + c T_{\as})$ with $ad-bc=1$.

This implies that I can then take $E'_{\AA} = \UEDA_{\AA}$, i.e. I can construct the deformed background using the original frame, such that:
\be
\gM'_{\MM\NN} = \UEDA_{\MM}{}^{\AA} \UEDA_{\NN}{}^{\BB} \mathcal{O}_{\AA}{}^{\CC} \mathcal{O}_{\BB}{}^{\DD} M_{\CC \DD}\,.
\ee
This can alternatively be viewed simply as a transformation of the generalised frame which preserves the algebra:
\be
\UEDA_{\MM}{}^{\AA} \rightarrow \UEDA'_{\MM}{}^{\AA} = \UEDA_{\MM}{}^{\BB} \mathcal{O}_{\BB}{}^{\AA} \,.
\ee
Either way, this then induces the transformations studied in the previous section.

In particular, the six-vector deformation corresponds to
\be
T_{\as}\rightarrow T_{\as}' = T_{\as} \,,\quad
T^{\as_1 \as_2 } \rightarrow T'^{\as_1 \as_2 }=T^{\as_1 \as_2 } \,,\quad
T^{\as_1 \dots \as_5 }\rightarrow T'^{\as_1 \dots \as_5} = T^{\as_1 \dots \as_5} + r^{\as_1 \dots \as_5 \bs} T_{\bs} \,,
\label{sixEDAdef}
\ee
preserving the $\Gsix$ EDA \eqref{sixEDA}.

\subsubsection*{Comparison with Yang-Baxter deformations}

The six-vector deformation with parameter \eqref{naturalSixVec} brings to mind existing approaches in the literature involving polyvector deformations.

A basic example of this sort is the (homogeneous) Yang-Baxter deformation \cite{Klimcik:2002zj,Klimcik:2008eq,Delduc:2013fga,Matsumoto:2015jja}, which has a particularly nice interpretation as an $\Odd$ transformation corresponding to a bivector deformation \cite{Sakamoto:2017cpu,Sakamoto:2018krs,Catal-Ozer:2019tmm}.\footnote{The use of manifest $\Odd$ symmetry is not the only way to identify this bivector, indeed it is implicit in the relationship between Yang-Baxter deformations and TsT transformations, see for instance \cite{Osten:2016dvf}, and has an alternative incarnation as the open string non-commutativity parameter appearing in the open/closed string map, as discussed in the Yang-Baxter context in \cite{vanTongeren:2015uha,vanTongeren:2016eeb} and also \cite{Araujo:2017jkb,Araujo:2017jap,Bakhmatov:2017joy,Borsato:2018idb}.}
I review this following \cite{Catal-Ozer:2019tmm, Sakatani:2019zrs}.
The starting point is a background admitting a (generically non-abelian) group isometry, with Killing vectors $k_a$ (corresponding to the right-invariant vectors of the group).
Let $r^{ab}$ be a constant antisymmetric matrix and define the following bivector:
\be
\beta^{(2)} = r^{ab} k_{a} k_{b} \,.
\label{ybbeta}
\ee
Then the bivector shift transformation of the generalised metric induces the Yang-Baxter deformation, if $r^{ab}$ obeys the classical Yang-Baxter equation.
This has a clean algebraic interpretation.
The $\Odd$ generalised metric describing the original background factorises as
\be
\cH_{MN} = E_M{}^A H_{AB} E_N{}^B \,,\quad 
E_M{}^A = \begin{pmatrix}
l_m{}^a & 0 \\ 
0 & v^m{}_a 
\end{pmatrix} 
\ee
The left-invariant vectors obey $[v_a, v_b] = - f_{ab}{}^c v_c$ and hence the generalised frames obey \eqref{genpar} in terms of an underlying Drinfeld double algebra of the form:
\be
\begin{split} 
[T_a, T_b]  = f_{ab}{}^c T_c \,,\quad
[T_a, \tilde T^b] = - f_{ac}{}^b \tilde T^c = - [\tilde T^b, T_a] \,,\quad
[\tilde T^a,\tilde T^b ]  = 0 \,.
\end{split} 
\ee
The bivector deformation corresponds to the following solution generating mechanism.
First carry out a constant bivector shift acting on the data that would appear after consistent truncation, in the same manner as \eqref{sixsgt}:
\be
H'_{AB} = \mathcal{O}_A{}^C \mathcal{O}_B^D H_{CD} \,,\quad
F'_{AB}{}^C = \mathcal{O}_A{}^D \mathcal{O}_B{}^E (\mathcal{O}^{-1})_F{}^C F_{DE}{}^F  
\,,\quad
\mathcal{O}_A{}^B \equiv \begin{pmatrix}
\delta_a^b & 0 \\
- r^{ab} & \delta^a_b 
\end{pmatrix} \,.
\ee
This corresponds to the following transformation of the generators of the Drinfeld double:
\be
T_a \rightarrow T'_a = T_a \,,\quad
\tilde T^a \rightarrow \tilde T'{}^a =  \tilde T^a - r^{ab} T_b \,.
\label{ybt} 
\ee
Assuming the homogeneous Yang-Baxter equation $f_{de}{}^{[a} r^{b|d}r^{|c]e} =0$, this leads to a new algebra 
\be
\begin{split} 
[T'_a, T'_b] & = f_{ab}{}^c T_c'\,,\quad
[T'_a, \tilde T'^b] = \tilde f^{bc}{}_a T'_c- f_{ac}{}^b \tilde T'^c = - [\tilde T'^b, T'_a] \,,\quad
[\tilde T'^a,\tilde T'^b ]  = \tilde f^{ab}{}_c \tilde T'^c \,,
\end{split} 
\ee
with $\tilde f^{ab}{}_c =  2 r^{d[a} f_{cd}{}^{b]}$.
A new generalised frame $E'_A$, of standard Poisson-Lie type, can then be constructed to realise this algebra, and is used to construct the new 10-dimensional deformed solution:
\be
\cH'_{MN} = E'_M{}^A H'_{AB} E'_N{}^B
 = E'_M{}^A \mathcal{O}_{A}{}^C E_C{}^P \cH_{PQ} E'_N{}^B \mathcal{O}_B{}^D E_D{}^Q  \,. 
\ee
The direct $\Odd$ transformation between $\cH'_{MN}$ and $\cH_{MN}$ is $\mathcal{O}_{M}{}^N = E'_M{}^A \mathcal{O}_{A}{}^BE_B{}^N$, and this is a non-constant bivector shift with parameter \eqref{ybbeta}.

This has been generalised from string to M-theory.
For backgrounds admitting at least $p$ Killing vectors, with $p=3$ or 6, polyvector deformations of the form $\Omega^{(p)} = r^{a_1 \dots a_p} k_{a_1} \dots k_{a_p}$ can be defined in terms of a constant antisymmetric tensor $r^{a_1 \dots a_p}$, which will obey a generalised Yang-Baxter equation. Examples of this type are discussed in \cite{Bakhmatov:2019dow,Bakhmatov:2020kul, Gubarev:2020ydf} and analysed from the EDA perspective in \cite{Sakatani:2019zrs,Malek:2019xrf,Malek:2020hpo,Musaev:2020nrt}.

The six-vector deformation \eqref{naturalSixVec} is distinguished from these deformations in two ways.
Firstly, it involves vectors which are not Killing.
Secondly, it does not modify the underlying algebra.
So while it is in the same ballpark as the (generalised) Yang-Baxter deformation, it is not exactly analogous to it.

A closer comparison can be made by thinking about possible extensions of these Yang-Baxter deformations such that the vectors involved are not Killing but are instead vectors associated with special non-isometric directions.
The simplest example would be to apply the $\Odd$ Yang-Baxter deformation \eqref{ybbeta} to a background of the type which is produced by non-abelian T-duality, meaning that it is characterised algebraically by non-vanishing dual structure constants, $\tilde f^{ab}{}_c \neq 0$, and vanishing $f_{ab}{}^c =0$. In this case I do not have to worry about the distinction between left- and right-invariant vector fields, and in \eqref{ybbeta} I can take $k_a =v_a$ where the $v_a$ are the commuting vectors generating translations in the non-abelian T-dual geometry.
These are again not symmetries of the spacetime, and are associated with the trivial `physical' isotropic subalgebra of the following Drinfeld double algebra:
\be
[T_a, T_b]  = 0 \,,\quad
[T_a, \tilde T^b] = \tilde f^{bc}{}_a T_c = - [\tilde T^b, T_a] \,,\quad
[\tilde T^a,\tilde T^b ]  = \tilde f^{ab}{}_c \tilde T^c  \,.
\ee
The transformation \eqref{ybt} produces the following algebra: 
\be
[T'_a, T'_b]  = 0 \,,\quad
[T'_a, \tilde T'^b] = \tilde f^{bc}{}_a T'_c = - [\tilde T'^b, T'_a] \,,\quad
[\tilde T'^a,\tilde T'^b ]  = \tilde f^{ab}{}_c \tilde T'^c - R^{abc} T'_c \,,
\ee
with
\be
R^{abc} = 3  \tilde f^{[ab}{}_d r^{c]d} \,.
\ee
This deformation therefore turns on `R-flux' structure constants, so that the algebra is no longer of the Drinfeld double type. 
This sort of algebraic deformation has appeared previously in the literature.
Normally it is required that the R-flux vanishes in order to stay within the confines of the Drinfeld double, for which the construction of associated generalised frames is known.
For example, the paper \cite{Lust:2018jsx} studied the permissible bivector transformations in a `non-abelian T-duality group', essentially corresponding to the set of $\Odd$ transformations preserving the splitting of the Drinfeld double into dual subalgebras. These transformations correspond to choices of $r^{ab}$ such that $R^{abc} = 0$.
This amounts to a cohomological condition (essentially requiring $r^{ab}$ be a cocycle).
A cohomological perspective is developed in detail in the analysis of \cite{Borsato:2021vfy} of possible $\Odd$ solution generating transformations (although the interest there is in transformations that act non-trivially on the generalised fluxes $F_{AB}{}^C$, rather than leave them invariant).

In my (fairly basic) example with $f_{ab}{}^c = 0$, if $R^{abc}=0$ then the bivector deformation identically preserves the structure constants of the Drinfeld double.
Then the original generalised frame can be used to describe the deformed geometry, $E'_A = E_A$.
This can alternatively be viewed simply as an algebra preserving redefinition of the generalised frame, $E'_M{}^A = E_M{}^B \mathcal{O}_B{}^A$, or as a direct $\Odd$ transformation between $\cH_{MN}$ and $\cH'_{MN}$ with parameter $\mathcal{O}_M{}^N = E_M{}^A \mathcal{O}_A{}^B E_B{}^N$. The latter is again a (not necessarily constant) bivector shift with $\beta^{mn} = r^{ab} v_a{}^m v_b{}^n$.
This is therefore directly analogous to the six-vector deformation \eqref{naturalSixVec}. 

An explicit example of this type, i.e. a choice of $r^{ab}$ which gives an \emph{invariance} of the algebra, is provided by:
\be
r^{ab} = \tilde f^{ab}{}_c n^c 
\label{rabn}
\ee
for some $n^c$. Then $R^{abc} = 0$ by the Jacobi identity for $\tilde f^{ab}{}_c$.

One reason why I am spending time on this example is that the deformation \eqref{rabn} is what is needed to view non-abelian T-duals, or Yang-Baxter deformed backgrounds, as T-folds.
The T-fold interpretation of such backgrounds was proposed in \cite{Fernandez-Melgarejo:2017oyu, Bugden:2019vlj}.
The main observation there was that for bivectors linear in the coordinates, a periodic identification of the coordinates required an identification of the background by a constant $\Odd$ bivector shift transformation.
Indeed, the bivectors appearing have the form $r^{ab} = \tilde f^{ab}{}_c x^c$ leading to \eqref{rabn} on identifying $x^c \sim x^c + n^c$.
This can again not be a standard T-duality as it does not act on isometric directions. 

This generalises to the U-fold interpretation of the 11-dimensional backgrounds studied in this paper (discussed already in \cite{Blair:2022gsx}).
This uses a generalised frame involving a trivector linear in the coordinates. 
Indeed, returning to the $\Gfour$ EDA \eqref{fiveEDA} with $f_{ab}{}^c = 0$, I can apply the following transformation, corresponding to a trivector shift
\be
T_a \rightarrow T'_a=T_a\,\quad
T^{ab} \rightarrow T'^{ab} = T^{ab} - r^{abc} T_c \,.
\ee
This leaves invariant the brackets involving $T'_a$, but produces
\be
[T'^{ab}, T'^{cd}]
= 2 \tilde f^{ab[c}{}_e T'^{d]e} + 2 \mathcal{R}^{[a,b]cde} T_e \,,\quad
\mathcal{R}^{a,bcde} = 4 r^{fa[b} \tilde f^{cde]}{}_f \,.
\ee 
Hence this is generically a non-trivial deformation of the EDA. The new structure constant $\mathcal{R}^{a,bcde}$ can be interpreted as an 11-dimensional R-flux \cite{Blair:2014zba}.
However, for certain parameters $r^{abc}$ the deformation vanishes (this presumably can be phrased in terms of three-algebra cohomology).
For example, a constant shift of the coordinates $x^i$ appearing in the generalised frame corresponds to taking
\be
r^{abc} = \tilde f^{abc}{}_d n^d \,.
\ee
As $2 \mathcal{R}^{[a,b]cde} = \tilde f^{cde}{}_f r^{abf} - 3 \tilde f^{ab[c}{}_f r^{de]f}$ (as can be checked using the Schouten identity) it follows that this then vanishes using the fundamental identity for three-algebras, namely $\tilde f^{abf}{}_g \tilde f^{cde}{}_f - 3 \tilde f^{ab[c}{}_f \tilde f^{de]f}{}_g=0$, which is obeyed as part of the consistency conditions of the EDA \cite{Sakatani:2019zrs,Malek:2019xrf}.

In summary, the six-vector deformation \eqref{naturalSixVec} generalises the form of a generalised Yang-Baxter deformation in supergravity, with (some of) the Killing vectors that normally appear in such a deformation being replaced by vectors associated with the underlying Poisson-Lie or Nambu-Lie group structure. 
This suggests that this deformation should be situated within a broader landscape of generalised duality transformations, or deformations, acting on backgrounds with and without isometries.
These deformations may correspond algebraically to turning on R-flux structure constants.
However as seen above, there are interesting examples where the deformation is an invariance of the algebra (and so will be cohomologically trivial in an appropriate sense). This is the case for the six-vector deformation \eqref{naturalSixVec}, which generates a non-trivial deformation of the geometry while leaving invariant the $\Gsix$ EDA \eqref{sixEDA}. This is also the case for the global $\Odd$ or $\Edd$ transformations which are needed to regard backgrounds involving a polyvector depending linearly on $d$ coordinates as T- or U-folds.

\section{Discussion}
\label{discussion}

In this paper, I developed further the generalised U-duality transformation \eqref{genUmap}, previously investigated in \cite{Blair:2020ndg,Blair:2022gsx}, mapping solutions of type IIA on $\mathrm{S}^3$ to new 11-dimensional solutions.
I provided general explicit formulae (in slightly simplified form in section \ref{generalmap} and in more detail in appendix \ref{appredup}) for how this map works at least restricted to the NSNS subsector on the initial IIA side.
This could be straightforwardly extended to include the RR sector, however this was not necessary for the main goal of this paper.

This main goal concerned the fate of TsT deformations of the type IIA solutions appearing in this generalised U-duality map.
It had been noted in \cite{Blair:2022gsx} that a TsT deformation of the solution with AdS${}_3 \times \text{T}^4 \times \mathrm{S}^3$ -- corresponding to undoing the near horizon limit of the F1-NS5 solution -- became a six-vector deformation in the new 11-dimensional solution.  
I reviewed this in section \ref{maptst}.
In section \ref{tstgeneral} I showed how this transmutation of TsT (or bivector) deformations into six-vector deformations is a generic feature of the solution generating mechanism \eqref{genUmap}.

The particularly interesting feature here is that the six-vector deformation involves an action of $\Gsix$ on isometric and non-isometric directions.
Hence it could also be thought of as a generalised U-duality transformation, with the special property being that it commutes with the trivector twist used in constructing the solution (this trivector twist introduces the coordinate dependence which breaks the isometries on which the $\Gsix$ U-duality group would otherwise act).
This six-deformation also preserves the form of the underlying exceptional Drinfeld algebra.

While the six-vector deformation was my main focus, I also briefly discussed how more general T-dualities result in trivector-dependent (and hence coordinate-dependent) $\Gsix$ transformations of the form \eqref{Utransmogrified}.
These could also be argued to be non-isometric or generalised U-dualities, with a less clear geometric interpretation than the six-vector deformation.
I focused on the $d=2$ case in this paper, but in general I would expect that $\Odd$ T-dualities of $\text{M}_7$ would map to $\mathrm{E}_{d+4(d+4)}$ transformations of the new 11-dimensional background.

In section \ref{algint}, I compared and contrasted the non-isometric six-vector deformation with (generalised) Yang-Baxter deformations.
I suggested that the six-vector deformation studied here should be thought of as part of a broader class of possible polyvector deformations, which can be defined in terms of a constant antisymmetric polyvector parameter and a set of vectors associated to `symmetries' of the solution.
Not all these symmetries need be isometries: the ones that are not isometries are `symmetries' in the sense that they are associated with the underlying Nambu-Lie group.
This intersects implicitly with previous observations on non-abelian/Yang-Baxter T-folds \cite{Fernandez-Melgarejo:2017oyu, Bugden:2019vlj} and touches on approaches to algebraically classify possible $\Odd$ deformations \cite{Lust:2018jsx, Borsato:2021vfy}. The latter presumably have a clear lift to the algebras appearing in generalised U-duality.

An initial application suggested by this perspective would be to find examples of duality deformations of non-abelian T-dual solutions involving the action of $\Odd$ or $\Edd$ transformations on the non-isometric part of the geometry. 
(Relatedly, it might also be possible to investigate whether non-abelian dualisations of $\text{M}_7$ survive in the 11-dimensional solution produced in \eqref{genUmap}.)

A potentially important point here is that as these transformations do not act on isometric directions, it is not immediately clear that they should be regarded as being on the same footing as conventional T-dualities acting on toroidal backgrounds with abelian isometries. The precise nature of the T-/U-fold interpretations mentioned above is therefore an interesting question.

This paper considered supergravity (and algebraic) aspects of generalised dualities only.
Their implications for brane or field theories is an intriguing direction for future work.

On the brane side, the expectation from the structure of the fluxes of 11-dimensional solution that IIA strings in $\text{M}_7$ should be identified with M5 branes wrapping the four-dimensional geometry $\Mfour$ in the new 11-dimensional setting.\footnote{This can also be checked by comparing the direct and double dimensional reductions \cite{BlairUpcoming}.}
So one can consider the question of whether or how properties of (deformations of) strings on type IIA are directly mapped to M5 branes in M-theory.
It is well-appreciated for instance that bivector shifts can be suggestively recast as open string non-commutativity parameters (as for example noted or used in the Yang-Baxter context in different ways in \cite{vanTongeren:2015uha,vanTongeren:2016eeb,Araujo:2017jkb,Araujo:2017jap,Borsato:2018idb}).
A corresponding interpretation of the six-vector deformation (and also the trivector twist) in terms of M-theory five-branes and membranes would be interesting.

On the field theory side, the obvious questions concern the original example of AdS${}_3 \times \text{T}^4 \times \text{S}^3$.
In section \ref{maptst} I noted that AdS-AdS and AdS-torus bivector deformations of AdS${}_3 \times \text{T}^4 \times \text{S}^3$  gave the same geometry after a simple coordinate shift. 
This is commented on in \cite{Chakraborty:2020xyz}, though there it is viewed as an accidental feature of the extremal solution.
It would be interesting to understand whether this has any deeper significance.
To this end I have included a short appendix \ref{genkill} describing the chain of dualities under which the AdS${}_3 \times \text{T}^4 \times \text{S}^3$ geometry is invariant. 
Now, the context for \cite{Chakraborty:2020xyz} is the study of the thermodynamic properties of the CFTs dual to the deformed spacetime geometry. 
This is of interest as TsT transformations of AdS${}_3 \times \text{T}^4 \times \text{S}^3$ correspond to (`single-trace') $T\bar T$ and $J \bar T$ deformations of the dual CFT${}_2$ \cite{Giveon:2017nie,Chakraborty:2019mdf,Apolo:2019zai} (the AdS-AdS deformation is $T \bar T$, the AdS-torus one is $J \bar T$ or $\bar J T$, and torus-torus example is $J \bar J$).
It is only natural to then speculate about the existence of an analogous deformation of the dual CFTs of the 11-dimensional AdS solutions constructed by the generalised U-duality \eqref{genUmap}.
These dual CFTs could be those analysed in \cite{Lozano:2020bxo}, which are dual to a possible global completion of the local solutions produced by \eqref{genUmap}. 
A starting point could be to analyse the thermodynamics of the non-extremal version of the new 11-dimensional solution, which was already written down in \cite{Blair:2022gsx}.

\section*{Acknowledgements}

I am supported by an FWO-Vlaanderen Postdoctoral Fellowship, as well as by the FWO-Vlaanderen through the project G006119N and by the Vrije Universiteit Brussel through the Strategic Research Program ``High-Energy Physics''. 
I would like to thank the following people for useful discussions and feedback on a draft of this paper: Alex Arvanitakis, Camille Eloy, Dan Thompson and Sofia Zhidkova.

\appendix

\section{Form and generalised geometry conventions} 
\label{conventions}
I define the Hodge dual of a $p$-form $F_{(p)}$ via
\be
( \star F)_{\mu_1 \dots \mu_{D-p} }  
= \tfrac{1}{p!} \sqrt{|g|}  \epsilon_{\mu_1 \dots \mu_{D-p} \nu_1 \dots \nu_p} g^{\nu_1\rho_1} \dots g^{\nu_p \rho_p} F_{\rho_1 \dots \rho_p}
\,.
\ee
Here $\epsilon_{\mu_1\dots \mu_D}$ denotes the Levi-Civita symbol $\epsilon_{01\dots D-1}= +1$.

The 11-dimensional 3-form, $C_{(3)}$, has field strength $F_{(4)} = \dd C_{(3)}$.
Its equation of motion is
\be
\dd \star F_{(4)} \rpm \tfrac12 F_{(4)} \wedge F_{(4)} = 0 \,.
\ee
I define $F_{(7)} = \star F_{(4)}$ 
and introduce the dual six-form via:
\be
\dd C_{(6)} = F_{(7)}  \rpm \tfrac12 C_{(3)} \wedge F_{(4)} \,.
\ee
The gauge transformations of these potentials are:
\be
\delta C_{(3)} = \dd \lambda_{(2)} \,,\quad
\delta C_{(6)} = \dd \lambda_{(5)} \rmp \tfrac12 \dd \lambda_{(2)} \wedge C_{(3)} \,.
\label{conventionsgauge}
\ee
Let $U=(u,\lambda_{(2)},\lambda_{(5)})$ and $V=(v,\omega_{(2)},\omega_{(5)})$ denote generalised vectors of $\Edd$ generalised geometry ($d\leq 6$) with generalised tangent bundle $\mathcal{E} \approx T\Mone \oplus \Lambda^2 T^*\Mone \oplus \Lambda^5 T^*M$.
Then the generalised Lie derivative is defined by:
\be
\mathcal{L}_U V = ( L_u v , L_u \omega_{(2)} - \iota_{v} d \lambda_{(2)} , L_u \omega_{(5)} - \iota_{v} d \lambda_{(5)} \rpm \omega_{(2)} \wedge d \lambda_{(2)} ) \,.
\label{genLie}
\ee
Comparing with \eqref{conventionsgauge} motivates decomposing the generalised vector $V$ as
\be
V = ( v , \bar\omega_{(2)} - \iota_v C_{(3)} , \bar\omega_{(5)} - \iota_v C_{(6)} \rpm C_{(3)} \wedge \bar \omega_{(2)} \rmp \tfrac12 \iota_v C_{(3)} \wedge C_{(3)} )  \,,
\ee
where $\bar V = (v, \bar\omega_{(2)}, \bar\omega_{(5)})$ is gauge invariant. 
(This geometric `twisting' is related to the definition of a choice of generalised frame \cite{Coimbra:2011ky} which naturally accomodates the gauge potentials on $\Mone$. The signs in \cite{Coimbra:2011ky} are slightly different as they take gauge parameters with opposite signs.)

In index notation, $V^{\MM} = (U^{-T})^{\MM}{}_{\NN} \bar V^{\NN}$ with
\be
(U^{-T})^{\MM}{}_{\NN}
 = 
\begin{pmatrix}
\delta^\ii_\jj & 0 & 0 \\
- C_{\ii_1 \ii_2 \jj} & 2 \delta^{[\jj_1}_{\ii_1} \delta^{\jj_2]}_{\ii_2} & 0 \\
- C_{\jj \ii_1 \dots \ii_5} \rmp 5 C_{\jj[\ii_1 \ii_2} C_{\ii_3 \ii_4\ii_5]} &
\rpm \tfrac{5!}{3!} C_{[\ii_1\ii_2\ii_3} \delta_{\ii_4}^{\jj_1} \delta_{\ii_5]}^{\jj_2} & 5! \delta^{[\jj_1}_{\ii_1} \dots \delta^{\jj_5]}_{\ii_5}
\end{pmatrix} \,.
\label{conventionsUtranspose}
\ee
From this I read off the matrix $U_{\MM}{}^{\NN}$ appearing in the factorisation of the generalised metric in terms of the $p$-form fields.
It is given by:
\be
U_{\MM}{}^{\NN} 
 = 
\begin{pmatrix}
\delta_{\ii}^{\jj} & C_{\ii\jj_1 \jj_2} & C_{\ii\jj_1 \dots \jj_5} \rmp 5 C_{\ii[\jj_1\jj_2} C_{\jj_3\jj_4\jj_5]} \\
0 & 2 \delta^{[\ii_1}_{\jj_1} \delta^{\ii_2]}_{\jj_2} &  \rmp \tfrac{5!}{3!} C_{[\jj_1\jj_2\jj_3} \delta_{\jj_4}^{\ii_1} \delta_{\jj_5]}^{\ii_2} \\ 
 0 & 0  & 5! \delta^{[\ii_1}_{\jj_1} \dots \delta^{\ii_5]}_{\jj_5}
\end{pmatrix} \,.
\label{conventionsU}
\ee
In order to describe the full 11-dimensional geometry, I need to make a split of the 11-dimensional spacetime as $\text{M}_{11-d} \times \Mone$.
Let $x^{\ii}$ denote $d$-dimensional coordinates on $\Mone$, and $x^\mu$ denote the remaining $11-d$ coordinates.
The 11-dimensional metric is decomposed as:
\be
ds_{11}^2  = |\phi|^{-\tfrac{1}{9-d}} g_{\mu\nu} dx^\mu dx^\nu + \phi_{\ii\jj} Dx^\ii Dx^\jj \,,\quad Dx^\ii \equiv \dd x^\ii + A_\mu{}^\ii dX^\mu \,,
\label{11Decomp}
\ee
where $\phi \equiv \det \phi_{\ii\jj}$. 
The three-form and its four-form field strength are decomposed as follows:
\be
 C_{(3)} = \Cdef_{(3)} + \Cdef_{(2) \ii } Dx^\ii  + \tfrac{1}{2} \Cdef_{(1) \ii\jj } Dx^\ii Dx^\jj + \tfrac{1}{3!} C_{\ii\jj\kk} Dx^\ii Dx^\jj Dx^\kk\,,
\label{Cdecomp_compact}
\ee
\be
  F_{(4)} = \Fdef_{(4)} + \Fdef_{(3) \ii } Dx^\ii  + \tfrac{1}{2} \Fdef_{(2) \ii\jj } Dx^\ii Dx^\jj + \tfrac{1}{3!} \Fdef_{(1) \ii\jj\kk} Dx^\ii Dx^\jj Dx^\kk + \tfrac{1}{4!} \Fdef_{\ii\jj\kk\ll} Dx^\ii Dx^\jj Dx^\kk Dx^\ll \,,
\label{Fdecomp_compact}
\ee
with implicit wedge products, and where the $(p)$ subscript denotes an $(11-d)$-dimensional $p$-form. 

U-duality mixes the components of the three-form and its dual.
Hence $C_{(6)}$ and $F_{(7)}$ are expanded similarly to \eqref{Cdecomp_compact} and \eqref{Fdecomp_compact}.

The $\Edd$ representations that result from the above decomposition have the following structure.
The external metric, $g_{\mu\nu}$, is an $\Edd$ singlet, invariant under $\Edd$ transformations.
The fields carrying purely $d$-dimensional indices enter a generalised metric parametrising a coset $\Edd/ H_{d}$ (with $H_{d}$ the maximal compact subgroup).
The remaining fields carrying $(11-d)$-dimensional indices are treated as components of $(11-d)$-dimensional forms, with each $p$-form gives a representation of $\Edd$. For instance, one has $\mathcal{A}_\mu{}^M \sim ( A_\mu{}^\ii, \Cdef_{\mu \ii\jj} ,\Cdef_{\mu \ii_1\dots \ii_5} ,\dots)$, modulo the inclusions of further exotic dualisations (only for $d\geq 7$) and further field redefinitions.
To avoid having to explain these redefinitions, I will more frequently work with the field strengths rather than the gauge fields themselves.
These field strengths can be determined using the fact that they transform under $d$-dimensional diffeomorphisms and gauge transformations as a generalised tensors.
In particular, this means $\mathcal{F}_{\mu\nu}{}^{\MM}$ factorises as $\mathcal{F}_{\mu\nu}{}^{\MM} = (U^{-T})^{\MM}{}_{\NN} \bar{\mathcal{F}}_{\mu\nu}{}^{\NN}$, with $\bar{\mathcal{F}}_{\mu\nu}{}^{\MM} = (F_{\mu\nu}{}^{\ii}, \Fdef_{\mu\nu \ii\jj}, \Fdef_{\mu\nu \ii_1 \dots \ii_5})$, using the matrix \eqref{conventionsUtranspose}.

\section{11-dimensional solutions on $\Mfour$ from type IIA on $\text{S}^3$} 

\label{appredup}

In this appendix I describe in more detail the derivation of the map \eqref{genUmap}, following the logic of \cite{Blair:2020ndg,Blair:2022gsx}.
This involves a reduction of type IIA on a three-sphere to $\mathrm{CSO}(4,0,1)$ gauged supergravity in seven dimensions, following by uplift to 11-dimensional supergravity on the space $\Mfour$.
I restrict to the NSNS subsector of type IIA on a three-sphere.
This leads to the expressions for the 11-dimensional solutions presented in section \ref{generalmap}.

\subsubsection*{Reduction: 10 to 7}

The starting point is the NSNS sector reduction ansatz for consistent truncation on $\text{S}^3$ \cite{Cvetic:2000dm,Cvetic:2000ah}.
Let $\mu^a$, $a=1,\dots,4$ be constrained coordinates on $\text{S}^3$, $\delta_{ab} \mu^a \mu^b= 1$.
I denote the seven-dimensional coordinates by $x^{\hI}$.
I introduce a unit determinant symmetric matrix
$M_{ab}$ with inverse $M^{ab}$, and define $U=2M^{ab} M^{bc} \mu^{a}\mu^{c}-\Delta M^{aa}$, $\Delta=M^{ab}\mu^{a}\mu^{b}$.
The metric, dilaton and three-form field strength are then written as:
\begin{equation}
\begin{split}
ds_{s}^{2} &=\Phi^{1/2}  g_{\hI \hJ} \dd x^{\hI} \dd x^{\hJ} +\frac{1}{g^{2}} \Delta^{-1}M_{ab}D\mu^{a}D\mu^{b}
\,,\quad
e^{2\varphi} =\Delta^{-1}\Phi^{5/4}\,,\\
H_{3} &=  \Fthree_{(3)}
- \tfrac{1}{2} \epsilon_{a_1 a_2 a_3 a_4}g^{-1}  \Delta^{-1}   F_{(2)}^{a_1 a_2} \wedge D \mu^{a_3} M^{a_4 b} \mu^b  
\\ & \quad
- \tfrac{1}{6}\epsilon_{a_1 a_2 a_3 a_4}g^{-2}  \Delta^{-2} \big(
U \mu^{a_1} D \mu^{a_2} \wedge D \mu^{a_3} \wedge D \mu^{a_4} 
+3 D\mu^{a_1} \wedge D \mu^{a_2}\wedge D M^{a_3 b} M^{a_4 c} \mu^b \mu^c 
\big)\,,
\label{S3ansatz}
\end{split}
\end{equation}
where $D \mu^a \equiv \dd\mu^a + g A_{(1)}{}^{ab} \mu^b$, $D M^{ab} = \dd M^{ab} + 2 g A_{(1)}{}^{(a|c} M^{c|b)}$ and $F_{(2)}{}^{ab} = \dd A_{(1)}{}^{ab} +g A_{(1)}{}^{ac} \wedge A_{(1)}{}^{cb}$. Numerical subscripts $(p)$ refer to form degree in seven dimensions.
The resulting $\Gfour$ covariant scalar matrix $\gM_{\fA \fB}$ and the two- and three-form field strengths are:
\be
\mathcal{M}_{\fA \fB} = \begin{pmatrix}
\Phi^{-\tfrac{1}{4}} M_{ab} & 0 \\
0 & \Phi
\end{pmatrix} \,,\quad
\Fa_{(2)}^{\fA \fB} = ( 0 , F_{(2)}^{ab} )\,,\quad
\quad
\Fb_{(3) \fA} = ( 0 ,\Fthree_{(3)}) \,.
\label{7data}
\ee
I can also work out the four-form $\mathcal{J}_{(4)}^{\fA}$, which is obtained via 
\be
\mathcal{J}_{(4)}^{\fA} = \rpm \star \mathcal{M}^{\fA \fB} \mathcal{H}_{(3) \fB} 
=  ( 0 , \rpm\Phi^{-1} \star_7 \Fthree_{(3)} ) \,.
\ee

\vspace{-3em}

\subsubsection*{Uplift: 7 to 11}

The 11-dimensional $\mathrm{CSO}(4,0,1)$  uplift of \cite{Blair:2020ndg,Blair:2022gsx} takes the form:
\be
\begin{split}
 g_{\hI \hJ}(x^{\hI},x^i) &=  g_{\hI \hJ}(x^{\hI}) \,,\quad
\gM_{\fM\fN}(x^{\hI},x^i)  =  E^{\fA}{}_{\fM}(x^i) E^{\fB}{}_{\fN}(x^i) \gM_{\fA\fB}(x^{\hI}) \,,\\
\Fa_{(2)}^{\fM \fN}(x^{\hI},x^i)  &= E^{\fM}{}_{\fA}(x^i)E^{\fN}{}_{\fB}(x^i) \Fa_{(2)}{}^{\fA \fB}(x^{\hI}) \,,\quad \Fb_{(3) \fM}(x^{\hI},x^i)   =   E^{\fA}{}_{\fM}(x^i)\Fb_{(3) \fA}(x^{\hI}) \,,\\
\Fc_{(4)}^{\fM}(x^{\hI}, x^i) & = E^{\fM}{}_{\fA}(x^i) \Fc_{(4)}^{\fA}(x^{\hI}) \,,
\end{split}
\label{ScherkSchwarz}
\ee
where all the fields on the left-hand side of these equalities are the $\Gfour$ covariant fields describing 11-dimensional supergravity in the exceptional geometric or exceptional field theory formalism \cite{Berman:2020tqn}.
This uplift uses an $\Gfour$ generalised vielbein represented here as a five-by-five matrix:
\be
 E^{\fA}{}_{\fM}(x^i) =\begin{pmatrix}
\delta^{a}_{m} & 0 \\
- g x_{m} & 1
\end{pmatrix}\,.
\label{CSOframe}
\ee
Using \eqref{ScherkSchwarz} with \eqref{7data} and \eqref{CSOframe} leads to
\begin{equation}
\mathcal{M}_{\fM \fN}
=\begin{pmatrix}
\Phi^{-\tfrac{1}{4}} M_{mn}+\Phi  g^2 M_{mp} M_{nq} x^p x^q & -\Phi g M_{mp} x^p \\
-\Phi g M_{np} x^p & \Phi
\end{pmatrix}\,,\quad
\end{equation}
\be
\Fa_{(2)}^{\fM \fN} = ( g x_j F_{(2)}^{ij} , F_{(2)}^{ij}  ) \,,\quad
\Fb_{(3) \fM} = ( - g x_m \Fthree_{(3)} , \Fthree_{(3)}  )\,, \quad
\mathcal{J}_{(4)}^{\fM} = ( 0,\rpm\Phi^{-1} \star_7 \Fthree_{(3)} ) \,,
\ee
where $F^{ij} \equiv \delta^i_a\delta^j_b F^{ab}$.

I now need the parametrisation of these fields in terms of the components of the usual 11-dimensional fields.
The $\Gfour$ generalised metric appears here in the five-dimensional representation with
\be
\gM_{\fM \fN} = (U_C)_{\fM}{}^{\fK} \GG_{\fK\fL} (U_C)_{\fN}{}^{\fL} \,,\,\, (U_C)_{\fM}{}^{\fN} = \begin{pmatrix}
\delta_i^j &0 \\ 
- C^j& 1 
\end{pmatrix} 
\,,\,\,
\GG_{\fM \fN} = \begin{pmatrix}
\phi^{-2/5} \phi_{ij} & 0 \\
0 & \phi^{3/5} 
\end{pmatrix} \,,
\label{5genmet2}
\ee
where $C^i \equiv \tfrac{1}{6} \epsilon^{ijkl} C_{jkl}$ and $\phi \equiv \det \phi_{ij}$.
The $\Gfour$ covariant field strengths can be identified with components of the 11-dimensional four-form and its seven-form dual as follows:
\be
\begin{split} 
\mathcal{F}_{(2)}{}^{i5}& = F_{(2)}{}^i\,,\quad\quad\,\,
\Fa_{(2)}{}^{ij} = \tfrac{1}{2} \epsilon^{ijkl} (\Fdef_{(2) kl} -  \Cdef_{klm} \Fdef_{(2)}{}^{m})\,,\\
\Fb_{(3) i} & = - \Fdef_{(3) i} \,,\quad
\Fb_{(3) 5}  =  \tfrac{1}{4!} \epsilon^{ijkl} (  \rpm \Fdef_{(3) ijkl} +  4  \Fdef_{(3) i}  \Cdef_{jkl})\,,\quad
\\
\Fc_{(4)}{}^5 & = - \Fdef_{(4)}\,,\quad\,\,
\Fc_{(4)}{}^i =  \tfrac{1}{3!} \epsilon^{ijkl}( \rmp \Fdef_{(4) jkl} - \Cdef_{jkl} \Fdef_{(4)} )
\label{deftildeF}\,.
\end{split} 
\ee
The signs are fixed in order that the $\Gfour$ covariant Bianchi identities match those of 11-dimensional supergravity.

I identify $g_{\hI \hJ}$ with the 7-dimensional $\Gfour$ invariant metric appearing in the decomposition \eqref{11Decomp} (on replacing the indices $\mu,\nu$ and $\ii,\jj$ there with $\hI,\hJ$ and $i,j$ here).
To reconstruct the full 11-dimensional metric, I also require the uplift of the one-form gauge potential:
\be
\Aa_{(1)}^{\fM \fN} = ( g x_j A_{(1)}^{ij} , A_{(1)}^{ij}  ) 
= ( A_{(1)}^i  , \tfrac12 \epsilon^{ijkl} \Cdef_{(1) kl} ) \,,
\label{Auplift}
\ee
where on the right-hand side I introduce the Kaluza-Klein vector and components of the three-form. Using the above expression for the $\Gfour$ generalised metric, the internal four-dimensional metric can be extracted.
Referring to the metric decomposition \eqref{11Decomp}, this leads to the full 11-dimensional metric:
\be
\begin{split} 
\dd s_{11}^2 & = 
 \left( \Phi^{-5/4} + g^2 M^{kl} x_k x_l \right)^{1/3} \Phi^{1/12}   g_{\hI \hJ} d x^{\hI} d x^{\hJ} 
\\ & \qquad + \frac{\left(  M_{ij} + \Phi^{5/4} g^2 M_{ip} M_{jq} x^p x^q \right)}{\left( \Phi^{-5/4} + g^2 M^{kl} x_k x_l \right)^{2/3} } \Phi^{-5/12}( \dd x^i + g x_r A_{(1)}^{ir} ) ( \dd x^j + g x_s A_{(1)}^{js} )\,.
\end{split}
\label{general11d}
\ee
The generalised metric also delivers the internal components of the three-form:
\be
C^i = \frac{g M^{ik} x_k}{ \Phi^{-5/4} +  g^2 M^{kl} x_k x_l} 
\Rightarrow 
C_{ijk} = \epsilon_{lijk}  \frac{g  M^{lm} x_m}{ \Phi^{-5/4} +  g^2 M^{pq} x_p x_q} \,.
\ee
From this and \eqref{Auplift}, two of the components of the four-form field strength in the decomposition \eqref{Fdecomp_compact} are determined, namely
\be
\Fdef_{ijkl} = 4 \partial_{[i} C_{jkl]} \,,\quad
\Fdef_{(1) ijk} = \dd_{(1)} C_{ijk} - L_{A_{(1)}} C_{ijk} - 3 \partial_{[i} \Cdef_{(1) jk]} \,.
\ee
Here $\dd_{(1)}$ means the exterior derivative with respect to the seven-dimensional coordinates only, and $L_{A_{(1)}}$ means the four-dimensional Lie derivative with respect to $A_{(1)}^i$.
These can be computed more explicitly, but the expressions are not immediately illuminating.
The dictionary for the field strengths, see \eqref{deftildeF}, then determines other components of both the four-form and its dual:
\be
\Fdef_{(2) ij} = \tfrac12 \epsilon_{ijkl} ( \mathcal{F}_{(2)}^{kl} + 2 C^k F_2^l ) 
= \tfrac12 \epsilon_{ijkl} (F_{(2)}^{kl}  + 2 C^k g x_m F_{(2)}^{lm})
\,,\quad
\Fdef_{(3) i} =  g x_i \Fthree_{(3)} \,.
\ee
\be
\begin{split} 
\Fdef_{(3) ijkl} 
= \rpm \epsilon_{ijkl} \Fthree_{(3)}  \frac{\Phi^{-5/4}}{\Phi^{-5/4} +  g^2 M^{kl} x_k x_l}\,.
\end{split} 
\ee
The latter can be dualised to obtain $\Fdef_{(4)}$. 
Alternatively, using $\Fc_{(4)}$, one finds
\be
\Fdef_{(4)} = \rmp \Phi^{-1} \star_7 \Fthree_{(3)} 
\ee
where $\star_7$ is defined using $ g_{\hI \hJ}$.
The full 11-dimensional four-form field strength then follows according to \eqref{Fdecomp_compact}.

On substituting $g_{\hI \hJ} = \Phi^{-1/2} G_{\hI\hJ}$, setting $M_{ab} =\delta_{ab}$ (which implies $e^{-2\varphi} = \Phi^{-5/4}$), $A_{(1)}{}^{ab} = 0$, and $g=1$ (for simplicity) I specialise to the case \eqref{origbackIIA} considered in the main text, leading to the 11-dimensional solution of equations \eqref{grand11metric} and \eqref{grand11Fs}.

\section{A self T-duality of AdS${}_3 \times \text{S}^1$}
\label{genkill}
Consider a background of the form:
\be
\dd s^2 = f ( - \dd t^2 + \dd x^2 )  + \dd y^2 \,,\quad
B= f \dd t \wedge \dd x \,,\quad
e^{-2\varphi} = 1 \,.
\label{M0}
\ee
I consider the following transformations:

\begin{itemize} 
\item Bivector shift with $\beta^{tx} = -\tfrac{\beta}{2}$. This produces the background:
\be
\dd s^2 = \frac{f}{1+\beta f} ( - \dd t^2 + \dd x^2 )  + \dd y^2 \,,\quad
B= \frac{f\dd t \wedge \dd x}{1+f\beta}  \,,\quad
e^{-2\varphi} = 1 + \beta f\,.
\label{M1}
\ee
\item Bivector shift with $\beta^{xy} = \gamma$. This produces the background:
\be
\dd s^2 = \frac{f}{1+f \gamma^2} ( - (\dd t + \gamma \dd y)^2 + \dd x^2) + \dd y^2 \,,\quad
B= \frac{f( \dd t + \gamma \dd y) \wedge \dd x}{1+\gamma^2 f}  \,,\quad
e^{-2\varphi} = 1+\gamma^2 f\,.
\label{M2}
\ee
\item Bivector shift with $\beta^{xy} = \gamma$ followed by a coordinate transformation with $t' =t+\gamma y$. This produces the background:
\be
\dd s^2 = \frac{f}{1+f \gamma^2} ( - \dd t'^2 + \dd x^2) + \dd y^2 \,,\quad
B= \frac{f\dd t' \wedge \dd x}{1+\gamma^2 f}  \,,\quad
e^{-2\varphi} = 1+\gamma^2 f\,.
\label{M3}
\ee
\end{itemize}
The backgrounds \eqref{M1} and \eqref{M3} agree on identifying $\beta=\gamma^2$.
As $U_\beta^{-1} = U_{-\beta}$, this implies the following transformation is an invariance of the original background \eqref{M0}:
\be
U_K \equiv U_{\beta_2} U_A U_{\beta_1} 
= \begin{pmatrix} 
A^{-T} & 0 \\ \beta_2 A^{-T} + A \beta_1 &  A
\end{pmatrix} \,,
\ee
where $U_A \equiv \text{diag} (A^{-T}, A) \in \mathrm{GL}(d) \subset \Odd$, with
\be
\beta_2 = \begin{pmatrix} 0 &  \frac12 \gamma^2 & 0 \\ -\frac12 \gamma^2 & 0 & 0 \\ 0 & 0 & 0 \end{pmatrix} \,,\quad
A = \begin{pmatrix} 1 & 0 & \gamma \\0 & 1 & 0 \\ 0 & 0 & 1 \end{pmatrix} \,,\quad
\beta_1 = \begin{pmatrix} 0 & 0 & 0 \\ 0 & 0 & \gamma \\ 0 & -\gamma & 0 \end{pmatrix} \,.
\ee
This could be interpreted as a generalised Killing isometry acting on doubled coordinates as
\be
\begin{array}{cccl}
t &\rightarrow t' &= &t + \gamma y - \tfrac{\gamma^2}{2} \tilde x\,,\\
x &\rightarrow x'& =&x + \gamma (\tilde y - \gamma \tilde t) + \tfrac{\gamma^2}{2} \tilde t\,,\\
y &\rightarrow y' &=& y - \gamma \tilde x
\end{array}\quad
\begin{array}{cccl}
\tilde t &\rightarrow \tilde t' &= &\tilde t\,,\\
\tilde x &\rightarrow \tilde x'& =& \tilde x\,,\\
\tilde y &\rightarrow \tilde y' &=& \tilde y- \gamma \tilde t \,.
\end{array}
\ee
The simplifed discussion above then embeds into the NSNS AdS${}_3 \times$ S$^3 \times$ T$^4$ solution, for which 
\be
\dd s^2 = f(r) ( - \dd t^2 + \dd x^2 )  + \frac{\dd r^2}{r^2}+ \sum_i  (\dd y^i)^2 + \dd s^2_{\text{S}^3} \,,\quad
B= f(r) \dd t \wedge \dd x +2 b \,,\quad
e^{-2\varphi} = 1 \,,
\label{MADS}
\ee
with $f(r) = r^2$.
The above shows that this background is invariant under: TsT on $(x,y^i)$ with parameter $\gamma$, coordinate shift on $(t,y)$ with parameter $\gamma$, TsT on $(t,x)$ with parameter $\frac{\gamma^2}{2}$.

\bibliography{CurrentBib}

\end{document}